\documentclass[fleqn,usenatbib]{mnras}

\usepackage{newtxtext,newtxmath}

\usepackage[T1]{fontenc}

\DeclareRobustCommand{\VAN}[3]{#2}
\let\VANthebibliography\thebibliography
\def\thebibliography{\DeclareRobustCommand{\VAN}[3]{##3}\VANthebibliography}

\usepackage{subfig}
\usepackage{graphicx}	
\usepackage{amsmath}

\usepackage{amssymb}	
\usepackage{caption}
\usepackage{xspace}

\newcommand{\rstar}{\ensuremath{R_{\star}}}
\newcommand{\bhac}{\texttt{BHAC}\xspace}

\title[Non-dipolar neutron star accretion in GRMHD]{GRMHD Simulations of Accreting Neutron Stars with Non-Dipole Fields}

\author[Das et al.]{
Pushpita Das,$^{1}$\thanks{E-mail:p.das2@uva.nl}
Oliver Porth,$^{1}$\thanks{E-mail:o.porth@uva.nl}
Anna L. Watts$^{1}$
\\
$^{1}$Anton Pannekoek Institute for Astronomy, University of Amsterdam, Science Park 904, 1098 XH, the Netherlands\\
}

\date{Accepted XXX. Received YYY; in original form ZZZ}

\pubyear{2021}

\begin{document}
\label{firstpage}
\pagerange{\pageref{firstpage}--\pageref{lastpage}}
\maketitle

\begin{abstract}
NASA’s NICER telescope has recently provided evidence for non-dipolar magnetic field structures in rotation-powered millisecond pulsars. These stars are assumed to have gone through a prolonged accretion spin-up phase, begging the question of what accretion flows onto stars with complex magnetic fields would look like. We present results from a suite of GRMHD simulations of accreting neutron stars for dipole, quadrupole, and quadrudipolar stellar field geometries. This is a first step towards simulating realistic hotspot shapes in a general relativistic framework to understand hotspot variability in accreting millisecond pulsars. We find that the location and size of the accretion columns resulting in hotspots changes significantly depending on initial stellar field strength and geometry. We also find that the strongest contributions to the stellar torque are from disk-connected fieldlines and the pulsar wind, leading to spin-down in almost all of the parameter regime explored here. We further analyze angular momentum transport in the accretion disk due to large scale magnetic stresses, turbulent stresses, wind- and compressible effects which we identify with convective motions. The disk collimates the initial open stellar flux forming jets. For dipoles, the disk-magnetosphere interaction can either enhance or reduce jet power compared to the isolated case. However for quadrupoles, the disk always leads to an enhanced net open flux making the jet power comparable to the dipolar case. We discuss our results in the context of observed neutron star jets and provide a viable mechanism to explain radio power both in the low- and high-magnetic field case.
\end{abstract}




\section{Introduction}

Many of the fundamental observational properties of neutron stars, such as their dynamics, emission, and thermal state, are associated with their strong magnetic fields.  However fundamental questions about those fields - their formation, structure and evolution - remain open \citep[for reviews see][]{Konar2017,Beskin2018,Igoshev2021}.  One such question is whether the magnetic field is a simple centered dipole, or whether neutron stars have offset or multipolar magnetic fields \citep[see e.g.][]{Jones1980,Barnard1982,HardingMuslimov2011fixed,Gralla2017,Sur2020,Petri2021}.  Theoretically this is certainly plausible:  higher order multipoles can be formed during the supernova \citep{Ardeljan2005,Obergaulinger2017}, or a complex field may develop during the neutron star's life as a result of processes like accretion \citep{Suvorov2020}.  

The NICER collaboration has recently provided strong evidence for the presence of multipolar magnetic field configurations in old, recycled rotation-powered millsecond pulsars \citep{Riley2019, Riley2021, Bilous2019, Miller2019, Miller2021}.   Offset dipole or quadrudipolar magnetic fields seem to be required to explain the location and shape of the magnetic poles.  It is then interesting to look at the evolutionary history of these stars, which are thought to have been through an extended accretion phase in a Low Mass X-ray Binary (LMXB) \citep{Alpar1982, Radhakrishnan1982,Bhattacharya1991}.  

Interaction between the accretion flow and the magnetic field is expected, and assumed to explain the phenomenon of accretion-powered millisecond pulsars (AMPs), where the bulk of the accretion flow is thought to be channeled onto the magnetic poles of the star.  Assuming that the star is not an aligned rotator, a clean view of the heated magnetic poles rotating in and out of the line of sight leads to rotationally-modulated X-ray emission.  However of the $\sim 130$ known neutron star LMXBs, there are only 20 AMPs (some of which are intermittent) \citep{PatrunoandWatts2012,DiSalvo20}.  This begs the question of why.  Perhaps the stars are aligned rotators \citep{Lamb2009}, or the accretion rate is high enough to overwhelm the magnetic field, disrupting channeled evolution  \citep{Psaltis1999b,Kulkarni2008,RomanovaKulkarni2008}.  Other options that have been discussed in the literature include suppression of pulsation amplitudes for high neutron star masses \citep{Ozel2009} or the effects of electron scattering, although the latter seems to have been ruled out \citep{Gogus2007}. The NICER results point to another possibility, which is that more complex field structures lead to less clearly channeled accretion, from which pulsations might not be visible. It is therefore interesting to consider the nature of the accretion flows that develop on stars with more complex fields.  We note that complex field topologies could be a generic feature of accreting stars, for example the magnetosphere of several  T-Tauri stars can be modeled with the addition of significant octopolar contributions  \citep{Donati2007,Donati2008}.  

 The nature of the flows that develop is also interesting for other reasons.  Accreting neutron stars exhibit many types of variability, the origin of which is not understood but is often attributed to interaction between the accretion flow and the star's magnetosphere \citep{Klis2006}.  The torques that arise in these flows regulate the spin distribution of the neutron star population, and determine whether or not neutron stars can be spun up to break-up frequency \citep[see for example][]{Andersson2005}.  Finally, a better understanding of the surface patterns that develop during accretion are important to efforts to use pulse profile modeling of accreting neutron stars to determine the dense matter equation of state \citep{Watts2016}.

We here perform the first axisymmetric general relativistic magnetohydrodynamic (GRMHD) simulations of accretion onto composite magnetic fields and contrast our findings with the standard dipolar case.  
Over the years, numerous simulations of accretion onto magnetized (neutron) stars have been carried out, both in the field of young-stellar-objects (YSO) and compact objects. However, the problem of MRI-driven accretion onto composite fields has not yet been studied by either community.

Non-relativistic simulations with $\alpha$-viscosity were presented for example by \cite{Romanova2002,Bessolaz2008,Fendt2009conf,Zanni2009,Zanni2013,RomanovaKoldobaEtAl2021}.  These have also been extended to include complex stellar field structures \citep{Long2007,Long2008} and approximations to the general relativistic potential of the compact object \citep{Kulkarni2005}.  
Yet another flavor of such simulations considers the turbulent viscosity driven by disk instabilities such as the magneto-rotational instability (MRI).  Non-relativistic 3D simulations of MRI turbulent disks were first presented in the seminal paper of \cite{Romanova2012} and more recently by  \citep{TakasaoTomidaEtAl2018} who have performed 3D simulations of accretion onto a weakly magnetized star, showing that fast funnel wall accretion streams can emerge even without stellar magnetosphere.  The first fully general relativistic 2D simulations of MRI turbulent disks were presented more recently by \citet[][hereafter PT17]{Parfrey2017}\defcitealias{Parfrey2017}{PT17}. %

In general, there are many similarities in the physics of star-disk interactions for compact objects and young stars.  For example, the process of disk-induced opening of the stellar magnetic flux and its effect on the spindown torque was discussed (among others) by \cite{Ferreira2000,MattPudritz2005a} for YSOs and by \cite{Parfrey2016} for millisecond pulsars.  Star-disk interactions have a strong impact on the accretion process and the relative strength of the magnetic field, the accretion ram pressure and rotation of the star is expected to give rise to discernible accretion states both for YSOs and neutron stars.  These are commonly described as ``boundary layer'', ``accretion column'' and ``propeller'' regimes (e.g. \cite{MattPudritz2005}, \citetalias{Parfrey2017}).  
Relativistic effects manifest themselves near the light-cylinder where the electric field becomes dynamically important.  Thus once the field-line rotation becomes relativistic, we expect a fourth regime where the Poynting flux due to the induced electric field makes a significant contribution to spin-down and energy extraction from the star which, if put to the extreme, can ablate the accretion disk via the ``pulsar wind'' (\cite{Parfrey2016}, \citetalias{Parfrey2017}).  While the pulsar wind is a truly relativistic effect, is worth mentioning that the additional spin-down contribution of a stellar wind was also considered in the context of T-Tauri stars by \cite{MattPudritz2005} and in the simulations of \cite{Zanni2009}. Hence while our study is focused on modeling accreting millisecond pulsars, as far as star-disk interactions and spin evolution are concerned, there are many points of intersection with the YSO case and we expect our results to have wider applicability.  

In this paper, we extend the work of \citetalias{Parfrey2017} to investigate GRMHD simulations with quadrupolar and quadrudipolar stellar fields.  In order to inform pulse-profile modeling, as a first step, we further study the structure and variability of the stellar hotspots that form in our simulations. Note that pulse profile modeling from compact stars crucially depends on spacetime curvature effects which leave a fingerprint on the pulse shape thus enabling to infer the compactness of the star \citep{PatrunoandWatts2012}.  This serves as strong motivation to perform our study in full general relativity.  
This paper is structured as follows: in Section \ref{sec:setup}, we describe our numerical setup, Section \ref{sec:results} elaborates on our results regarding accretion states and overall morphology of the flow, spin-down torques, jet formation and accretion hotspots.  We discuss our results and conclude in Section \ref{sec:discussion}.  
While in principle star-disk interactions can be studied within a newtonian framework the strong spacetime curvature at the stellar surface leaves a fingerprint on the emission from surface hotspots which asks for a general relativistic treatment \citep{PatrunoandWatts2012}.  

\section{Numerical Setup}\label{sec:setup}

We solve the GRMHD equations using \bhac \citep{Porth2017, Olivares2019} in Schwarzschild coordinates  
\begin{align}
\nabla_{\mu}(\rho u^{\mu}) &= 0 \\
\nabla_{\mu} T^{\mu\nu} &= 0 \\
\nabla_{\mu} ^{\star}F^{\mu\nu} &= 0 \, .
\end{align}
Here $T^{\mu\nu}$ is the energy-momentum tensor of an ideal magneto-fluid; $^{\star}F^{\mu\nu}$ is the dual of the Faraday tensor $F^{\mu\nu}$ and $\rho$,  $u^{\mu}$, are rest-mass density and fluid 4-velocity respectively. 

Our inner boundary of the domain is set to the stellar radius, $\rstar = 4r_g$ (${\rm r_g = GM/c^2}$). In the magnetosphere ($r<r_{\rm LC} = c/\Omega$), the initial pressure and density is set such that magnetization ($\sigma = b^2/\rho$) and plasma beta ($\beta = 2p/b^2$) are 100 and 0.01 respectively and smoothly transitions to follow $r^{-6}$ outside the light cylinder. We initialize the domain with a  standard hydrodynamic equilibrium  \cite{Fishbone1976}- torus with constant angular momentum, inner edge r$_{\rm in} = 60r_g$ and density maximum located at $85r_g$. The torus is magnetised with poloidal loops defined by $A_{\phi} \propto {\rm max(\rho/\rho_{\rm max} - 0.2,0)}$ such that $2p_{\rm max}/b^2_{\rm max}=100$ (where the subscript max indicates the individual maxima of the quantities as customary in black hole torus setups \cite[e.g.][]{Porth2019}). Outside the torus, the magnetosphere is initialised with the stationary, axisymmetric vector potentials for the vacuum multipole field in Schwarzschild space-time following \cite{Wasserman1983}. We perturb the pressure with $10\%$ white noise in order to excite the magnetorotational instability (MRI) inside the torus.  At the start of the simulations, the stellar rotation is switched on by prescribing $v^\phi=\Omega$ at the inner radial boundary.  

Regarding the boundary conditions, the problem presents three distinct physical regions at $r=\rstar$: (a) the wind region, (b) the accretion column region and (c) the closed field deadzone.  Our choice of numerical boundary conditions follows physical considerations in each of these regions.  
We extrapolate pressure and density in the accretion columns and set them to initial (low) values in the non-accreting regions. 
In particular the boundary for the magnetic field requires special attention:  everywhere at $r=\rstar$, we fix $B^{\theta}$ to the value given by the initial vector potential and let $B^r$ follow from demanding $\nabla \cdot \mathbf{B}=0$ in our staggered discretization of the magnetic field.  
Regarding the toroidal field component, we assign $B^{\phi} = 0$ in the closed zone and use extrapolation in wind- and column- regions. In practice, this is realized by dynamically tracing out the stellar magnetic fieldlines every few timesteps via the particle module of \bhac.  Here, all the fieldlines returning back to the star are tagged as closed and all the open fieldlines or ones ending up in the disk are identified accordingly.  
This is an important distinction as without specifying $B^\phi$ in the deadzone, the boundary would be underdetermined meaning that the deadzone solution would depend on the details of the  extrapolation, e.g. leading to spurious $B^\phi\ne 0$.  The need for providing one additional boundary condition in the deadzone becomes apparent as by definition, the deadzone fieldlines do not cross the light-cylinder which otherwise supplies one internal boundary condition (see e.g. \citealt{Bogovalov1997fixed,Porth2010} and \cite{McKinney2006c} for the boundary conditions in force-free simulations of aligned rotators).\footnote{On the other hand, we have verified that fixing globally $B^\phi=0$ leads to a discontinuity in the wind zone Poynting flux which is under-estimated by a factor of $\approx3$ compared to the analytic pulsar wind solution.}  
To provide velocity boundary conditions for both the accreting and force-free regimes, we project the velocity onto the magnetic field direction and extrapolate the parallel velocity component $v_{\|}$ following \citetalias{Parfrey2017}.  Rotation is enforced by adding a constant $\Omega$ in the azimuthal direction.  Hence in the force-free regions we essentially enforce uniform rotation with $\Omega$ while in the accretion columns we allow matter to leave the domain along the field lines. We verified our force-free boundary conditions against the aligned rotator solution \citep[e.g.][]{Gruzinov1999,McKinney2006c} and recover the expected spin-down power with an agreement of $<5\%$.

We assume that the parts of the magnetosphere disconnected from the disk (the wind and deadzone regions) are filled with tenuous plasma resulting in high magnetization similar to the isolated pulsar.  In the context of T-Tauri stars, it has been argued that some fraction of the accreted mass is can be diverted from the stellar surface into a stellar wind \citep{MattPudritz2005}. Whether this is possible for neutron stars depends on how mass and energy spreads from the accretion hotspot and thus relies on modeling of the surface physics which is beyond the scope of this work.  Generally though it is thought that the accreted mass forms an ``ocean'' from which material can be lifted up only during brief episodes of thermonuclear bursts  \cite{GallowayKeek2021,KeekArzoumanianEtAl2018}.  In line with this, we here assume that the magnetization in the wind- and deadzone regions is ``high'', rendering the dynamics in these regions close to the time-dependent, force-free, degenerate electrodynamic limit of relativistic MHD \citep[][]{Komissarov2002}.    
 
 To maintain force-free like conditions in the magnetosphere ($r<r_{\rm lc}$) we adopt the prescription of \cite{Tchekhovskoy2013} and \citetalias{Parfrey2017} hence we drive pressure (p), density ($\rho$) and $v_{\|}$ to target values $\rho_{\rm t} = b^2/\sigma_{\rm t}$, $p_{\rm t} = b^2 \beta_{\rm t}/2$ and $v_{\|,{\rm t}} = 0$. The variables $k = (\rho,p,v_{\|}$) are driven towards $k_{\rm t} = (\rho_{\rm t},p_{\rm t},u_{\|,{\rm t}}$) with the prescription $k = k_{\rm t} + (k - k_{\rm t}){\rm exp}(-\kappa\Delta t/\tau)$ at the end of each time-step following \cite{Tchekhovskoy2013}. Here ${\rm \kappa = (cos\theta_m,cos\theta_m,1)}$, ${\rm \theta_m}$ = magnetic colatitude, $\Delta t$ is the time-step and $\tau=\tau(r)$ is the driving time-scale.  We adopt immediate driving ($\tau$ = 0)  within $r< 0.5r_{\rm lc}$ and $\tau(r)$ is chosen to smoothly switch off driving ($\tau \rightarrow \infty$) for $r\to r_{\rm lc}$ such that the MHD solution will never be modified beyond $r_{\rm lc}$.  
 
 In order to distinguish the accreting fluid from the force-free regions, we introduce a tracer-fluid ($\mathcal{T}$)  with the torus initialized as $\mathcal{T} = 1$ and the rest of the domain with $\mathcal{T} = 0$. The tracer is passively advected with the flow by solving the extra equation $\nabla_{\mu}({\mathcal{T}\rho u^{\mu}}) = 0$. The tracer hence separates the regions of the flow: in the magnetosphere, thus for $\mathcal{T} = 0$, we drive $\rho$,p and ${\rm v_{\|}}$ to target values as discussed above. Inside the accretion flow ($\mathcal{T} = 1$), the solution follows that of the unmodified ideal GRMHD equations. In the transition region $0<\mathcal{T}<1$, the solution is a mixture of force-free and MHD as we interpolate with weight $\mathcal{T}$ between the two solutions.  

 For numerical stability, next to the driving described in the previous paragraph, we maintain global floor and ceiling values for $\beta$ and $\sigma$ respectively, and inject gas pressure and density such that $\beta > 10^{-3}$ and $\sigma < 10^3$  respectively (common practice in simulations of magnetically arrested black hole disks, Olivares et al. in preparation). In practice, this is only activated in the early evolutionary times of the dipolar cases where magnetic field is compressed against the incoming accretion flow.  
 
 For the time evolution, we use a total variation diminishing Lax-Friedrich scheme for fluxes along with Piecewise Parabolic reconstruction scheme and second order modified Euler time-stepper. The domain is logarithmically spaced in the radial direction extending from $r_{\rm in} = 4r_g$ to $r_{\rm out} = 30000r_g$.  This ensures that the outer boundaries are causally disconnected from the region of interest for the entire duration of the simulations.  
 We employ a three-level adaptive mesh (AMR) with base resolution of $N_{\rm r} \times N_{\theta} = 480\times192$ cells, resulting in an effective numerical resolution of $1920 \times 768$ cells.  

\subsection{Units and fiducial scaling}\label{subsec:units}
In the following sections we report simulation quantities in code units where we have $c=1, r_g=1$.  
Since our simulations neglect the effect of radiation (which is a reasonable assumption for $L_{\rm Acc}\lesssim 1\% L_{\rm Edd}$), once the stellar radius is fixed,
we have the freedom to constrain one additional variable to scale our simulations to a physical system.  For definiteness we set the radius  $R_\star=10\rm km$.  
Choosing a fiducial polar field strength of $B_{0,\rm cgs}$, e.g. $10^8\rm G$ \

means that all quantities of interest can be converted to their cgs values via multiplication with the corresponding scale factors.  For a simulation with a given dipolar moment ($\mu$)\footnote{This scaling also applies for quadrupole topologies with the substitution $\mu=Q/2.76$ where Q is the quadrupolar moment.}, the conversion factors are
\begin{align}
    \rho_{\rm cgs} &= 10^{-6} \left(\frac{B_{0,\rm cgs}}{10^8 G}\right)^2 \left(\frac{\mu}{30}\right)^{-2}\rm g\, cm^{-3} \\
    \mu_{\rm cgs} &= 1.6\times 10^{24} \left(\frac{B_{0,\rm cgs}}{10^8 G}\right) \left(\frac{\mu}{30}\right)^{-1}\rm G\, cm^3 \\
    \dot{M}_{\rm cgs} & = 3\times 10^{-11} \left(\frac{B_{0,\rm cgs}}{10^8 G}\right)^2 \left(\frac{\mu}{30}\right)^{-2} M_{\odot}\, \rm year^{-1}
\end{align}

Since the polar field strength implicitly sets the scaling for the mass accretion rate, it is worthwhile checking up to which polar field strength the assumption of radiatively inefficient flow remains valid.  For an accretion rate in code units $\dot{M}$ (typically $0.1-0.4$, see Table \ref{tab:runparameters}), the Eddington factor becomes 
\begin{align}
L_{\rm Acc}/L_{\rm Edd} = 2\times 10^{-3} \left(\frac{B_{0,\rm cgs}}{10^8 G}\right)^2 \left(\frac{\mu}{30}\right)^{-2} \dot{M}
\end{align}
where the accretion luminosity is simply defined as $L_{\rm Acc} := G M \dot{M} /R_{\star}$.  Hence the radiatively inefficient accretion flow (RIAF) assumption is expected to hold up to polar field strengths of $B_{0,\rm cgs}\simeq 2.4\times 10^{8} \mu_{30}\,\dot{M}^{-1/2}\rm G\approx 10^9 \rm G$.  
With these choices, the scaling for the rotation of the star becomes
\begin{align}
    P = 1.75 \left(\frac{\Omega}{0.03}\right)^{-1}\rm ms \, .
\end{align}

\section{Results}\label{sec:results}
We perform a set of runs for different parameters and different initial stellar magnetic field topology as summarized in Table \ref{tab:runparameters}.  The table also reports the averages of accretion rate and torque extracted at the stellar surface.  In general, our simulations give rise to highly variable accretion as quantified by the variability index $c_{\rm v}\sim 1-5$ that compares the standard deviation with the mean of $\dot{M}$ (see also Appendix \ref{sec:mdot} for  time-series of the accretion rates). 
We will first discuss general solution characteristics for dipoles (\ref{sec:dipoles}), quadru(di)poles (\ref{sec:quadrupoles}) and thereafter quantify the accretion torques (\ref{sec:torque}) and outflows (\ref{sec:jets}).  We will return to the issue of variability in Section \ref{sec:hotspots} on the accretion hotspots.  

\begin{table}
 \setlength{\tabcolsep}{0.35\tabcolsep}
 \caption{Parameters for different models. $\mu$, Q, $\dot{M}$ and $\dot{L}$ are given in code units with $\dot{M}$, $\dot{L}$ extracted at $r = \rstar$. Angular brackets ($\langle \rangle$) denote the average of the quantities over time for t $\in$ [10000, 30000]$r_g/c$. ${\rm c_v} =$ Standard deviation($\dot{M}$)/mean($\dot{M})$ shows the variability index. $r_{\rm m,num}$ shows the approximate range of magnetospheric radius as seen in the simulations and $r_{\rm m}$ shows the theoretical magnetospheric radius ($\xi r_{\rm A}, \xi = 0.7$) using $\dot{M}$ extracted at the stellar surface in the dipolar case. $B_{\rm di}(\rstar,0)/B_{\rm t}(\rstar,0)$ shows the dipolar fraction at $\theta = 0$ for quadrudipolar fields at the stellar surface.} \label{tab:runparameters}
\centering
 \begin{tabular}{*{8}{c}} 
 \hline
 Field Strength & $\Omega$ & $\langle\dot{M}\rangle$ & $\langle\dot{L}\rangle$ & ${\rm c_v}$ & $r_{\rm m,num}$ & $r_{\rm m}$\\ [0.5ex] 
 \hline
Dipoles $(\mu)$ & $[c/r_g]$ & & & $\frac{{\rm SD}(\dot{M})}{\langle\dot{M}\rangle}$ &[$r_g$] & [$r_g$]\\
 \hline
5 & 0.03  & 0.2361 & 0.7246 & 1.0253 & $\rstar(=4)$ & 2.4\\
30 & 0.03  & 0.2124 & 1.3141 & 1.4765 & [6.5,10] & 6.8\\ 
45 & 0.03  & 0.1845 & 2.1344 & 2.5822 & [12, 15]& 9.0\\
70 &  0.03 & 0.1157 & 2.9979 & 2.8489 & [13,15]& 13.3\\ 
160 &  0.03 & 0.0855 & 6.4696 & 4.0206 & [19.5,22]& 23.2\\ 
250 &  0.03 & 0.0526 & 11.5847 & 5.1989 & [21,23]& 34.4\\ 
30  &  0.016 & 0.4615 &  0.3713 & 1.0126 & [6.5,10] & 5.5\\
30  &  0.02 & 0.4269 & 0.5851 & 0.9844 & [6.5,10] & 5.6\\
30  &  0.04 & 0.1566 & 2.0605 & 2.2646 & [6.5,10] & 7.4\\
30  &  0.05 & 0.1112 & 2.2218 & 2.9187 & [6.5,10] & 8.2\\
\hline
 Quadrupoles (Q) & &\\
\hline
25 & 0.03  & 0.2613 & 0.2880 & 0.9852\\ 
70 & 0.03  & 0.2734 & 0.3015 & 0.8605\\
82.8529 & 0.03  & 0.298 & 0.3323 & 0.9369\\
120 & 0.03  & 0.2199 & 0.3116 & 1.1707\\
250 & 0.03  & 0.1089 & 0.209 & 2.1647\\
82.8529 & 0.016  & 0.3096 & 0.0009 & 0.8236\\
82.8529 & 0.02  & 0.2552 & 0.1469 & 0.9211\\
82.8529 & 0.04  & 0.2717 & 0.5180 & 1.0978\\
82.8529 & 0.05  & 0.2343 & 0.9245 & 1.3433\\
350 & 0.1  & 0.004 & 13.6665  &  \\ 
\hline 
 Quadrudipoles ($\mu$, Q) & & & & & $\frac{B_{\rm di}(\rstar,0)}{B_{\rm t}(\rstar,0)}$\\
\hline
$\mu$ = 25, Q = 13.8088 & 0.03  & 0.2508 & 1.1077 & 1.395 & 0.8333\\
$\mu$ = 20, Q = 27.6176 & 0.03  & 0.2587 & 1.0323 & 1.0662 & 0.6667\\
$\mu$ = 15, Q = 41.4264 & 0.03  & 0.301 & 0.8272 & 1.2141 & 0.5000\\
$\mu$ = 10, Q = 55.2352 & 0.03  & 0.2378 & 0.4902 & 1.0539 & 0.3333\\
$\mu$ = 5,  Q = 69.0441 & 0.03  & 0.2744 & 0.5225 & 1.0241 & 0.1667\\
\hline
\end{tabular}
\end{table}
\subsection{Dipoles}\label{sec:dipoles}
For reference we discuss the evolution for three different accretion states for stars with dipolar magnetic fields.  Here we adopt a fiducial value of $\Omega=0.03$ and vary the magnetic moment $\mu\in\{5,30,160\}$. The general time evolution is described as follows: as we switch on stellar rotation at t = 0, the rotating star launches an Alfv\'en wave which engulfs the torus. As the wave propagates outwards, within a few lightcrossing times across the lightcylinder ($r_{\rm lc} = 1/\Omega$), the inner regions of the magnetosphere relax to the steady-state solution of the isolated pulsar \citep{komissarov2006, Spitkovsky2006}. At around $\approx 2800 GM/c^3$, the MRI driven turbulence starts to drive material inwards. This transient phase lasts until $\sim 10\, 000 GM/c^3$, afterwards we obtain a quasi-stationary state which is characterized by a constant average accretion rate and angular momentum flux and as well as constant open magnetic flux from the star (see e.g. Figures \ref{fig:stress1d}, \ref{fig:openfluxvstime}).  
In the equatorial plane, material proceeds until a balance between stellar magnetic pressure and ram pressure of the accreting material is obtained, defining the magnetospheric radius $r_{\rm m}$.  
Customarily one defines the Alfv\'en radius
\begin{align}
    r_{\rm A} = \bigg(\frac{\mu^4}{2GM\dot{M}^2}\bigg)^{1/7}
    \label{eq:rm}
\end{align}
\citep{Elsner1977} and sets $r_{\rm m} = \xi r_{\rm A}$ with $\xi\le 1$.  Typically $\xi\simeq0.5$ \citep[e.g.][]{Bessolaz2008,Zanni2013}.  In our simulations, we find that the magnetospheric radius is well captured by the slightly larger parameter $\xi\approx 0.7$, which might be explained by the fact that our thick disk simulations are closer to being spherical, as assumed in the classic derivation of the Alfv\'en radius.  The ranges of the simulated $r_{\rm m, num}$ are compared with the analytic expectation for $\xi=0.7$ in Table \ref{tab:runparameters}.

\begin{figure*}
   \centering
     \includegraphics[width=17cm, angle=0]{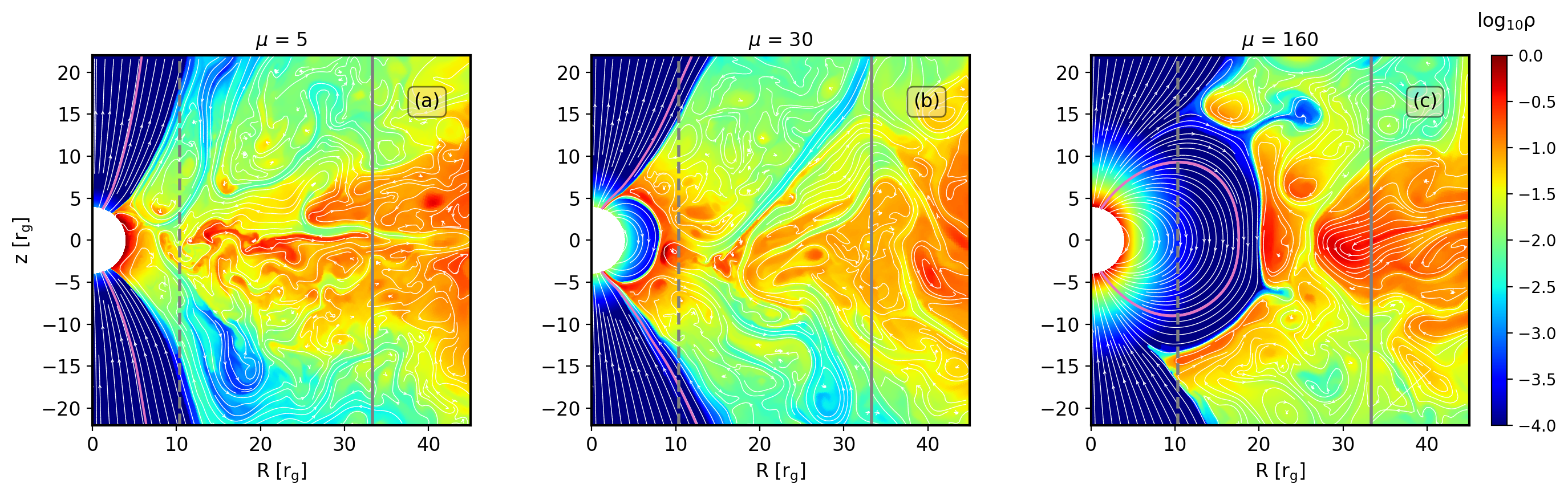}%
     \caption{Different accretion regimes depending upon magnetic strength of the star for angular frequency of $\Omega=0.03$ in the quasi stationary state at $t = 19000r_g/c$. (a) Boundary layer, (b) channeled accretion and (c) Propeller regime. The solid and dashed lines represent the light cylinder radius and the corotation radius respectively. The pink lines show the last closed fieldline for respective isolated pulsars.}%
    \label{fig:diffmudipole}
    \vspace{-4mm}
\end{figure*}

Depending on $\mu$, the location of the magnetospheric radius will give rise to various accretion states as described by \citetalias{Parfrey2017}. The logarithmic densities along with the characteristic radii are illustrated in Figure \ref{fig:diffmudipole}.
In the first panel (\ref{fig:diffmudipole}a), we recover the \textit{boundary layer regime} with $r_{\rm m}<R_{\star}$ for $\mu=5$.  
Increasing $\mu$ leads to higher $r_m$ where the disk couples to the magnetosphere to form accretion columns yielding the \textit{channeled accretion regime} (Figure \ref{fig:diffmudipole}b). 
A further increase in $\mu$ results in the \textit{propeller regime} with $r_{\rm co} < r_{\rm m} < r_{\rm lc}$ (Figure \ref{fig:diffmudipole}c), where 
\begin{align}
    r_{\rm co} = (r_{\rm lc}^2 r_g)^{1/3} \label{eq:corotation}
\end{align} 
is the expected corotation radius for a disk in Keplerian rotation. However, as we will discuss below, the actual corotation radius in the simulations is somewhat smaller due to the sub-Keplerian nature of the inner disk (see e.g. Figure \ref{fig:omega}). 
In this state, the magnetosphere centrifugally ejects most the accreting gas from the corotation radius while allowing a few occasional streams to reach the stellar surface.  
In the propeller regime (at high values of $\mu=160,250$), Table \ref{tab:runparameters} shows that Eq. (\ref{eq:rm}) predicts larger magnetospheric radii than observed numerically.  This can be explained by the fact that in this regime, a significant amount of mass is diverted into an outflow before reaching the stellar surface.  The theoretical $r_{\rm m}$ values however use the lower $\dot{M}$ extracted at the stellar surface which results in an over-estimation of the true $r_{\rm m}$.
\subsection{Quadrupoles and quadrudipoles}\label{sec:quadrupoles}
\begin{figure*}
   \centering
    \subfloat{\includegraphics[width=17cm]{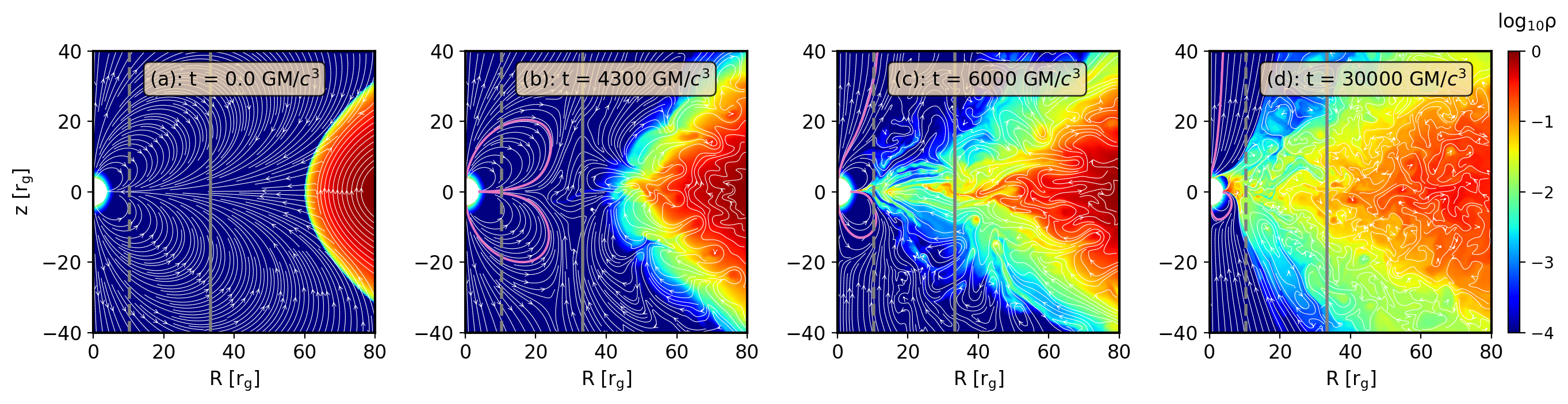}}%
    \caption{Evolution of the disk in presence of a pure quadrupole with Q = 70, rotating at $\Omega=0.03$. The different panels represent logarithmic density profiles at different snapshots. All the labels are same as in Figure \ref{fig:diffmudipole}.}%
    \label{fig:Q70time-evolu}
\end{figure*}
\begin{figure}
   \centering
    \includegraphics[width=8.5cm, angle=0]{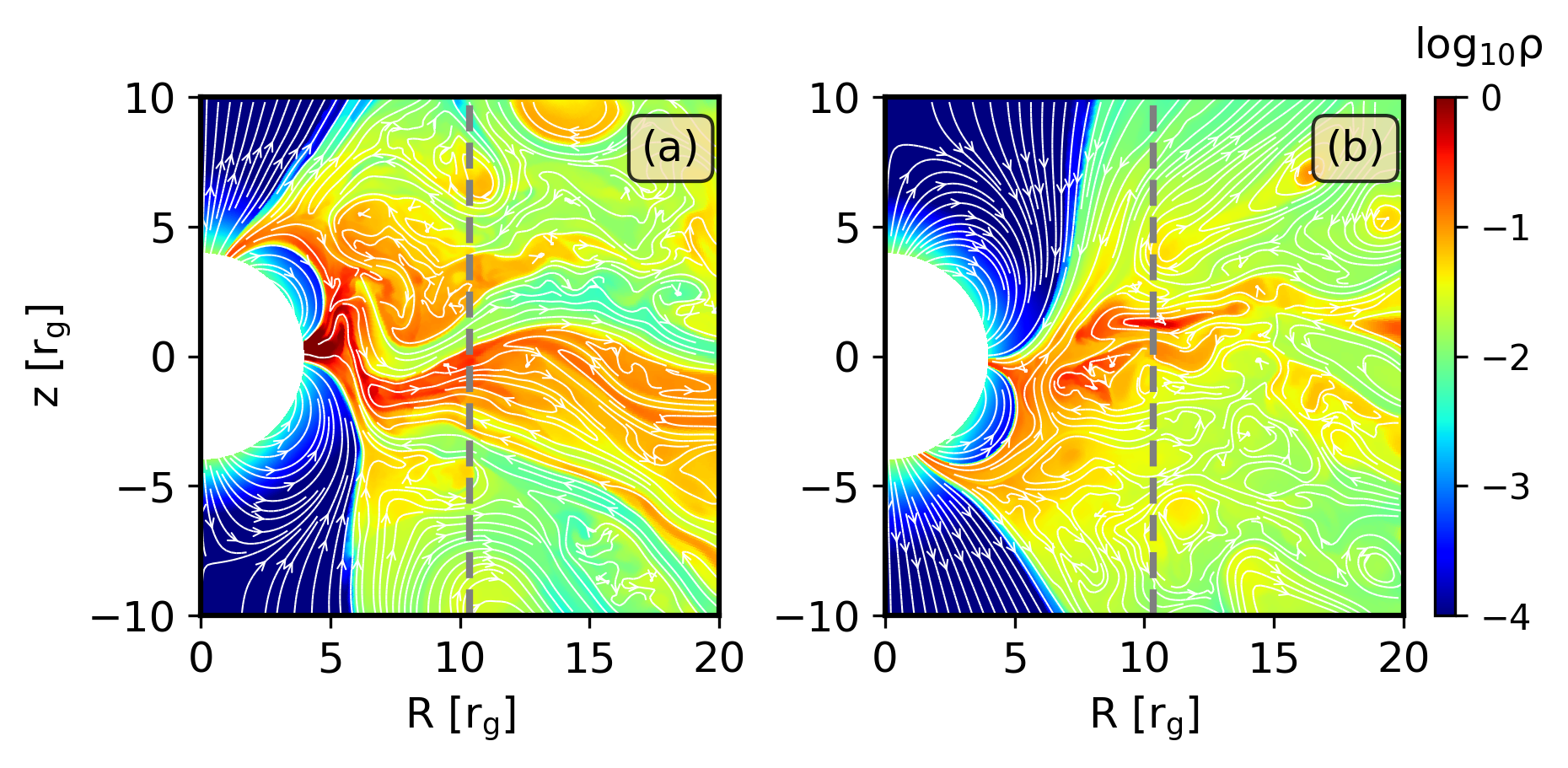}
    \caption{Different hotspot locations for different star-disk magnetic field orientation for Q = 70, $\Omega=0.03$ at t = $25000 r_g/c$. Here the left panel shows density profile ($\log_{10}(\rho)$) for initial disk magnetic field anti-parallel to the stellar field in the northern hemisphere and parallel in the southern hemisphere (our default configuration). The opposite case is shown in the right panel. All the labels are same as in Figure \ref{fig:diffmudipole}.}%
    \label{fig:diff-disk}
    \vspace{-4mm}
 \end{figure}

Next, we initialize the star with a quadrupolar magnetic field. 
As in the dipolar case, field lines initially open up at the the light cylinder radius, however, in the isolated quadrupolar case two radial current sheets are obtained at angles of roughly $\arctan(2)\simeq 63^\circ$ from the poles. The time evolution is illustrated in Figure \ref{fig:Q70time-evolu} for Q = 70 where the initial disk magnetic field is anti parallel to both the stellar closed zone and the equatorial open flux in the northern hemisphere, and parallel in the southern hemisphere. As the gas travels inward, the disk magnetic field reconnects with the stellar field and eventually opens up the previously closed stellar flux in the northern hemisphere. Meanwhile the field in the lower hemisphere is compressed, shifting the lower y-point towards the pole. The asymmetric disk-magnetosphere coupling modifies the initial quadrupolar magnetosphere to resemble a quadrudipolar field in the vicinity of the stellar surface. Once inside the light-cylinder, the gas flows primarily through the quadrupolar opening at the equator, forming a ``belt''. For our simulation with Q = 70 , $\Omega$ = 0.03, we hence obtain two accretion columns, one at the equator and other closer to the northern pole \footnote{\url{https://youtu.be/58CXMXrmhy8}}. The location of the polar hotspot depends on the initial star-disk magnetic field configuration which is further illustrated in Figure \ref{fig:diff-disk} where the initial disk magnetic field is flipped. 
The open equatorial flux which contributes to the southern jet shields the lower closed zone, thus preventing formation of a southern hotspot (Figure \ref{fig:Q70time-evolu}d). However, this highly asymmetric inner magnetospheric configuration seems to have only a small impact on the net open flux in the upper and lower hemispheres resulting in relatively symmetric jets (see section \ref{sec:jets} for further discussion).  
 
 To investigate the dependence of the accretion column structure as function of quadrupole moment, we perform two sets of experiments in the regime $r_{\rm m}< r_{\rm co}$: in the first sweep, we adopt a pure quadrupole with increasing  moment Q$\in\{25,70,120,250\}$.  In the second sweep, for a given polar field strength, we successively increase the quadrupole contribution from the pure dipole case to the pure quadrupole case.  
We opt to fix the total polar field strength in the second sweep since this parameter can in principle be derived from observations of cyclotron resonant scattering features \citep[see][]{StaubertTrumperEtAl2019} which are thought to constrain the polar field strength (but note \citet{KylafisTrumperEtAl2021a} for a discussion of the caveats with the probed location).  This allows us to investigate how the accretion process changes for a given polar field strength (determined model-independently) when changing the dipole to quadrupole ratio. 
 \begin{figure*}
    \centering
    \subfloat{\includegraphics[width=17cm]{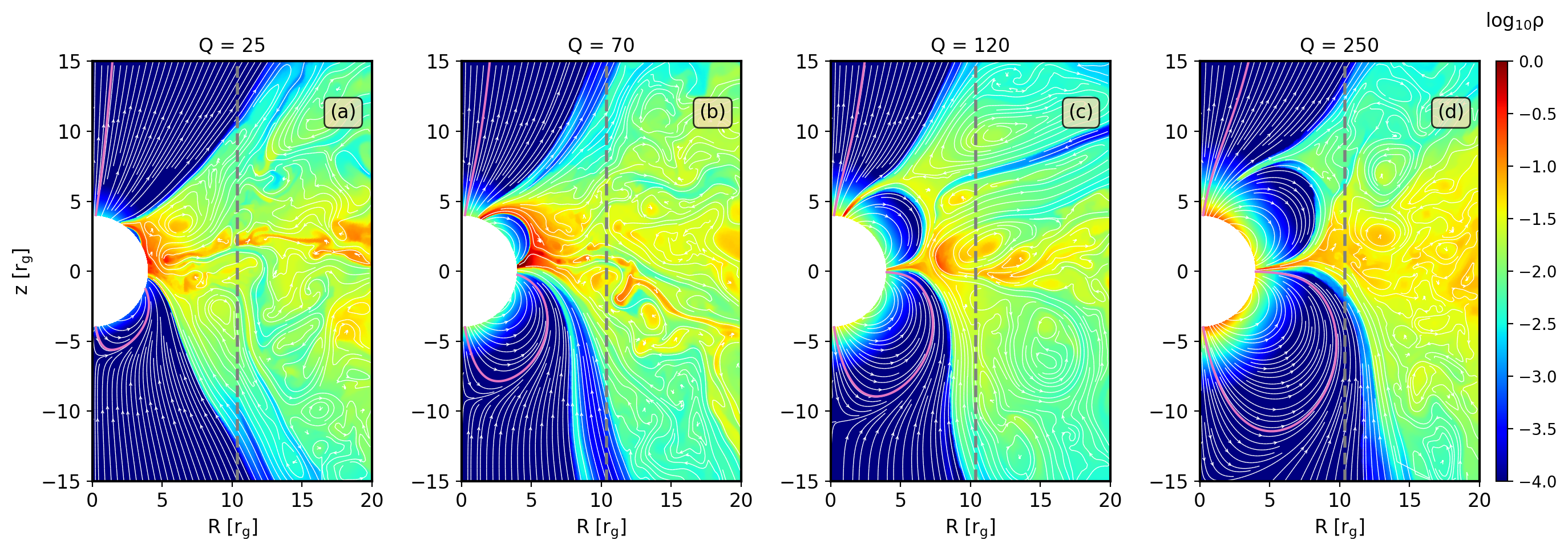}}
    \caption{Different hotspot size for different magnetic strengths. Density profiles at t = $28120 r_g/c$. (a) Q = 25, (b) Q = 70, (c) Q = 120, (d) Q = 250. All the labels are same as in Figure \ref{fig:diffmudipole}.}
    \label{fig:diffQ}
\end{figure*}
 Figure \ref{fig:diffQ} shows density profiles in the quasi-stationary state for different quadrupolar strengths in the first set. 
 At the lowest quadrupolar moment (Q = 25), the disk crushes the northern closed zone leading to boundary layer accretion through upper hemisphere. Increasing quadrupolar strength (Q) has two main effects: 
 first, it halts the disk at increasingly larger radii as in the dipolar case. 
 Second, it affects the width of the accretion columns. Truncation of the disk at larger radii leads to thin accretion columns in the equatorial region.  
 There is a pronounced asymmetry of the upper and lower deadzone with the lower one being more extended. This can be attributed to differential disk induced flux opening which is discussed in more detail in Section \ref{sec:jets}. 
 
 \begin{figure*}
   \centering
    \subfloat{\includegraphics[width=17cm]{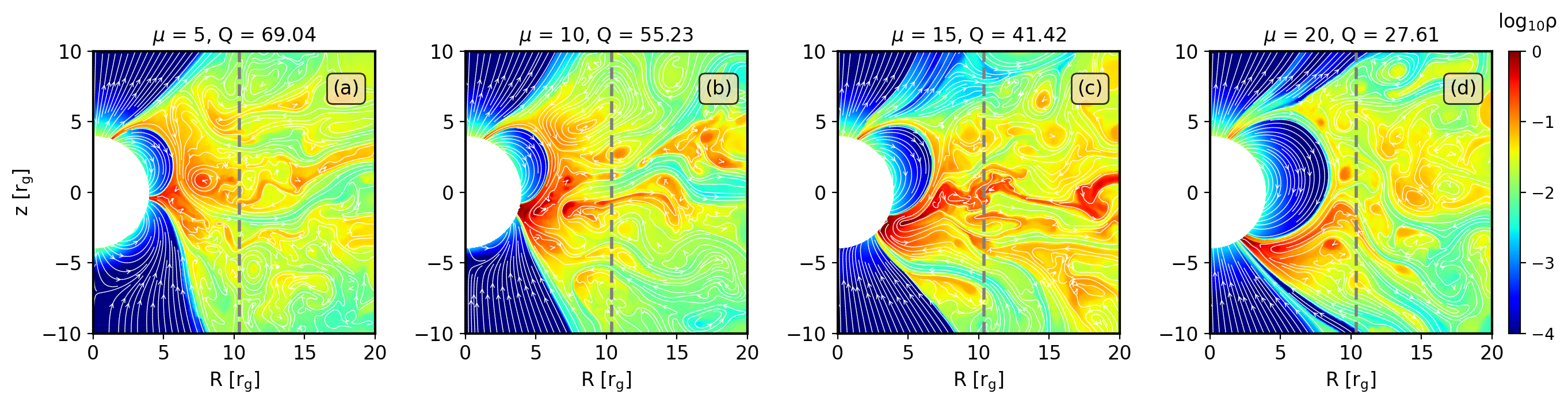} }%
    \caption{Effect of increasing quadrupolar contribution on the accretion flow (right to left). Different panels represent the logarithmic density profiles at t = $20270 r_g/c$ for different dipolar and quadrupolar contribution, where Q and $\mu$ are varied such that the surface polar field strength stays constant. All the labels are same as in Figure \ref{fig:diffmudipole}.}
    \label{fig:diffQ-mu}
    \vspace{-4mm}
 \end{figure*}
  The non-equatorial accretion stream also grows weaker with the increase in magnetic strength. 
 For the highest quadrupolar strength in our case (Figure \ref{fig:diffQ}(d)), the upper accretion column has become so weak that it becomes difficult to resolve numerically and often stops midway between the corotation radius and the stellar surface. 
 The stopping of the upper column is however not a purely numerical effect and is expected from physical grounds: for one, 
 we observe that a strong asymmetric pulsar wind ablates column material to entrain it along the outflow.  Further, as $r_{\rm m}$ approaches $r_{\rm co}$, imperfect disk-magnetospheric coupling occurs when a blob of gas, instead of coupling to the stellar fields at $r_{\rm m}$ retains its Keplerian velocity. Often, such blobs of gas ``hovering'' on top of the column within $r_{\rm co}$  either get ejected along the jet or accreted through the equatorial hotspot. This effect is more prominent in the quadrudipoles. 

Figure \ref{fig:diffQ-mu} shows the distribution of matter in presence of a quadrudipolar stellar magnetic field. While in the pure quadrupole, increasing Q had the main effect of increasing the magnetospheric radius while leaving the column footpoints roughly in place; for the quadrudipoles, increasing the quadrupolar contribution leads to a progressive shift towards the northern hemisphere.  
Further, since the magnetic pressure of the quadrupole decreases more rapidly, the magnetospheric radius also moves inwards with an increasing quadrupolar contribution.  This is best seen in the diminishing of the closed zone with increasing Q in Figure \ref{fig:diffQ-mu}.  
\subsection{Torque}\label{sec:torque}
The accretion torque is directly extracted from the angular momentum flux through the stellar surface defined as
\begin{align}
    \dot{L} = 2\pi\int_0^\pi T^r_{\phi} \sqrt{-g}d\theta \, .\label{eq:ldot}
\end{align}
The torque experienced by accreting stars can be divided into two components. The ``matter'' torque due to accreting material and the ``EM'' torque due to the stress exerted by the electromagnetic fields. The corresponding tensorial components read
\begin{align}
{\text{T}^{\text{MA}}}^{r}_{\phi} &= (\rho_0 + u_g + p_g)u^{r}u_{\phi}  \\
{\text{T}^{\text{EM}}}^r_{\phi} &= b^2u^{r}u_{\phi} - b^{r}b_{\phi}
\label{angularmomentum}
\end{align}
Here $u_g$ is the internal energy density, $p_g = (\hat{\gamma} - 1)u_g$ is the ideal gas pressure, $\hat{\gamma}$ is the adiabatic index, $p_b = b^{\mu}b_{\mu}/2 = b^2/2$ is the magnetic pressure and $b^{\mu}$ describes the fluid frame magnetic field as defined for example in \cite{Porth2017}.
\begin{figure*}
   \centering
    \subfloat{\includegraphics[width=18cm, angle=0]{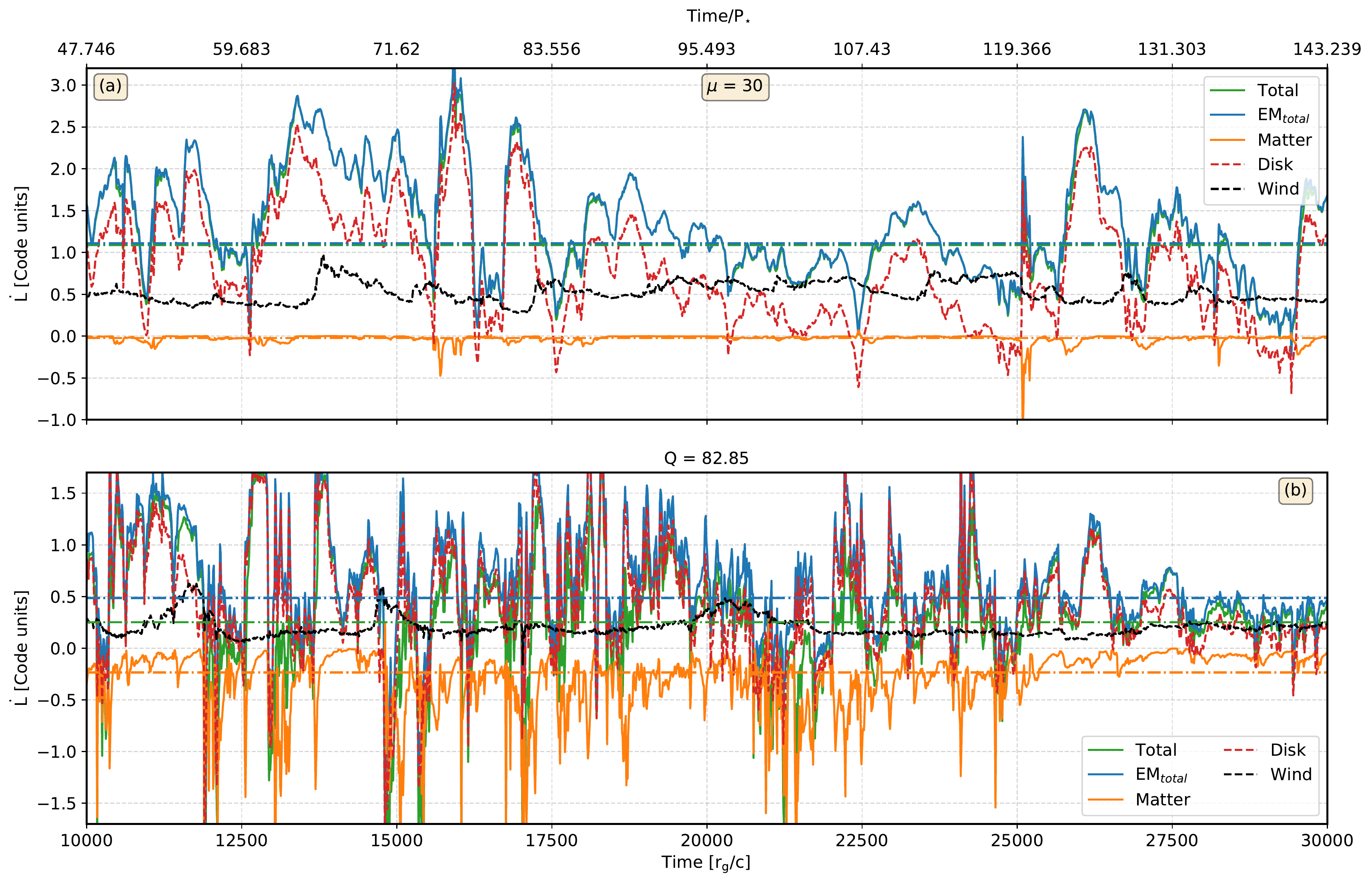}}
    \caption{Torque evolution of the neutron star with a dipolar field, magnetic moment with $\mu = 30$ (upper panel) and quadrupolar field with Q = 82.85 (lower panel) rotating at an angular velocity of $\Omega=0.03$, both having same polar magnetic field strength at the stellar surface. Here the green solid line shows the time evolution of total torque and blue and orange solid lines show the total electromagnetic (EM) and matter torque respectively. The total EM torque is further separated into wind and disk contribution where the black and red dashed lines show the wind torque and the `disk connected' torque (see Section \ref{sec:torque}) respectively. The blue, orange and green dash-dotted lines represent the averaged values ($t \in [10000,30000]r_g/c$) for EM, matter and the total torque respectively. The upper x-axis denotes time in terms of stellar period with $P_{star} = 209.43 r_g/c$.}%
    \label{fig:torque_dipole}
\end{figure*}

Figure \ref{fig:torque_dipole} illustrates the evolution of angular momentum flux at the stellar surface for $\mu = 30$ and Q = 82.85 for $\Omega$ = 0.03. As a positive angular momentum flux means that angular momentum is transported away from the star, a positive (negative) torque therefore leads to spin down (up) of the star. The upper panel in Figure \ref{fig:torque_dipole} represents the evolution of the total torque for a dipolar field at the stellar surface. For stars having magnetic strength high enough to couple to the flow outside of the star ($r_{\rm m} > \rstar$), at $r_{\rm m}$, the accreting gas transfers most of its angular momentum to the electromagnetic fields via magnetic stress and correspondingly, matter contributes only a negligible torque (\citet{Romanova2002}; \citetalias{Parfrey2017}) at the stellar surface. The remaining matter torque is always negative, thus acting towards spin-up.
 
The electromagnetic torque can be either positive or negative depending upon $r_{\rm m}$ and $r_{\rm co}$ and can be further divided into two components, an enhanced spin-down torque from the pulsar wind \citep{Parfrey2016}, and a variable spin up/down torque from the stellar fieldlines connected to the disk. We separate these two components by considering whether the corresponding fieldlines are ``open'' or ``disk-connected''.  We define an open field line as one that reaches past the light cylinder (in practice, we demand $r > 2r_{\rm lc}$ or $r \sin\theta > r_{\rm lc}$) and stays within the highly magnetized polar region characterized by $\sigma>1$.  All other field lines are either closed dead-zone field or attach to the accretion flow via the column and outflow impacting on the disk. The fieldlines connected to the disk inside and outside of $r_{\rm co}$ result in spin-up and spin-down respectively. In our simulations, the dominant spin-down component comes from the disk-connected fieldlines (Figure  \ref{fig:torque_dipole}), however, the wind torque contributes on a similar level.  

The lower panel of Figure \ref{fig:torque_dipole} shows the time evolution of torque at the stellar surface for an initial quadrupolar stellar magnetic field. For these parameters, the magnitude of the total torque averaged over time (t $\in$ [10000, 30000]$r_g/c$) is roughly half for the quadrupolar case (both cases effectively spin down the star). Several factors come into play that tend to weaken the torque in the quadrupolar case: (a) For quadrupoles (Q = 82.85), since the disk can come much further in towards the stellar surface, the fraction of fieldlines inside the corotation radius is larger compared to dipoles. Thus, the spin-down torque from the ``disk-connected'' field is smaller compared to dipoles.  
(b) For quadrupolar fields, most of the time, matter can directly hit the stellar surface through the equatorial belt. This leads to a larger spin-up matter torque component (around 40 percent of the total torque) compared to dipoles. 
(c) Lastly, also the wind torque is roughly $2$ times smaller in the quadrupolar case with Q=82.85 (see also Section \ref{sec:jets}).  
All these factors contributes to less spin-down for quadrupoles as compared to dipoles for same polar magnetic field strength. 

\begin{figure}
   \centering
    \subfloat{\includegraphics[width=7cm]{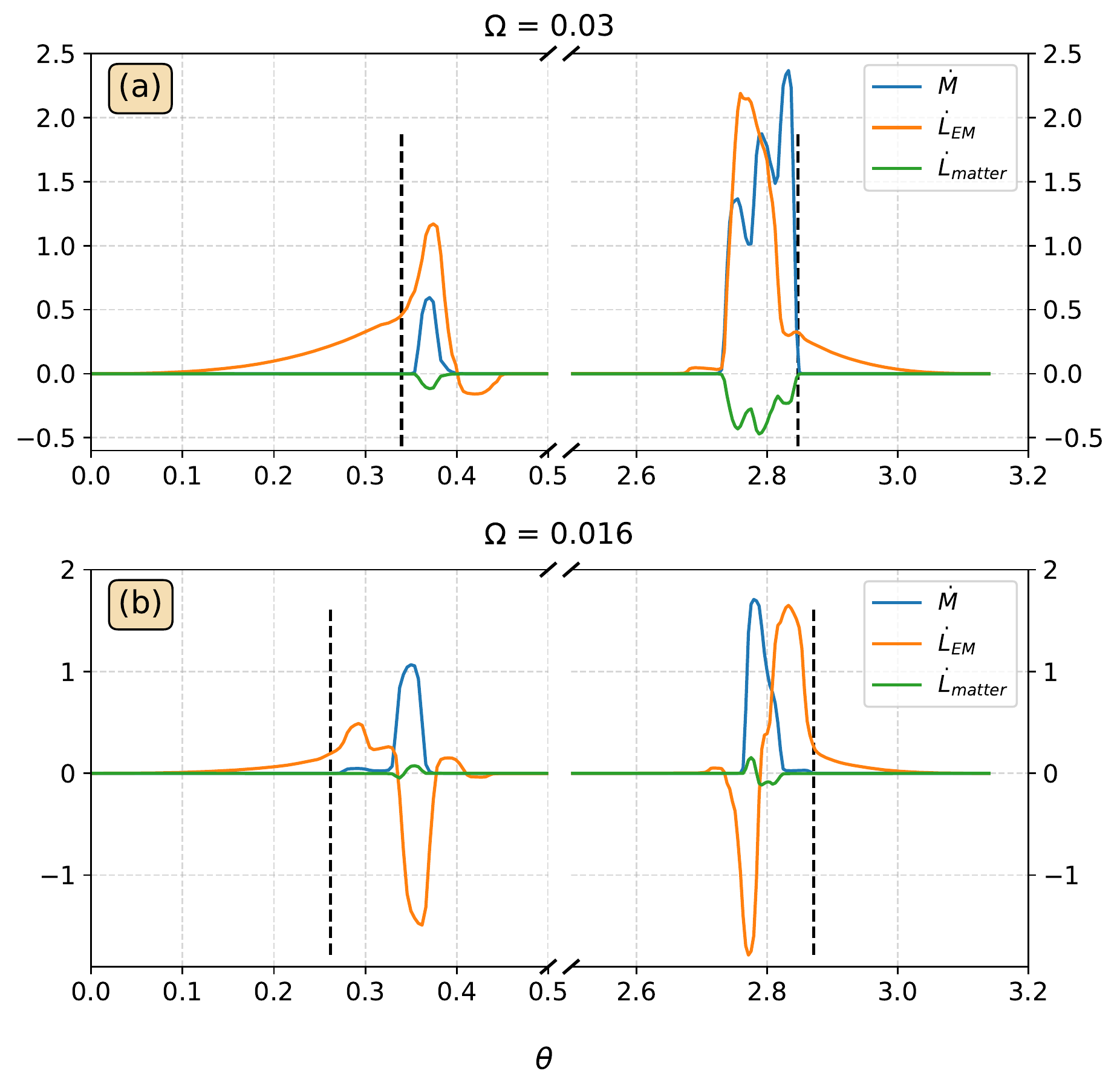}}%
    \caption{Different torque components as a function of $\theta$ at the stellar surface for a dipolar field with $\mu$ = 30 (a) $\Omega = 0.03$ at t = 25790$r_g/c$, (b) $\Omega = 0.0166$, at t = 26000$r_g/c$. The black dashed line represents the extent of the wind zone from the poles in both hemispheres (defined by angular positions of fieldlines reaching either $2r_{\text{lc}}$ or $r \sin\theta > r_{\text{lc}}$ while ensuring $\sigma > 1$).}
    \label{fig:mu30-om0p0166-om0p03}
\end{figure}

Figure \ref{fig:mu30-om0p0166-om0p03} shows the instantaneous angular profiles of the  torque at the stellar surface for a dipolar field with $\mu$ = 30 and $\Omega=0.03$ (top) and $\Omega=0.016$ (bottom). The black dashed lines illustrate the separation of the wind zone from the ``disk-connected'' zone as per our criteria.  One can see that at this point, the electromagnetic torque departs from the wind profile near the poles and increases sharply in the accretion column, also characterized by non-vanishing $\dot{M}$ and spin-up matter torques.   \\
As we decrease $\Omega$ to $0.016$, the scale separation between $r_{\rm m}$ and $r_{\rm co}$ increases and the electromagnetic stresses in the column start contributing to spin-up.  This is best seen in the northern column (around $\theta\simeq0.35$ in Figure \ref{fig:mu30-om0p0166-om0p03}).  
However, overall, the average total torque for $\Omega = 0.016$ is still positive, spinning down the star. 
According to Eq. (\ref{eq:corotation}), the classical corotation radius for $\Omega = 0.016$ is $\simeq 16r_g$ which is roughly a factor two larger than the magnetospheric radius for the run in question.  One might hence wonder why we don't observe a stronger spin-up effect due to the magnetic stresses of the column.  

 \begin{figure}
   \centering
    \includegraphics[width=7cm]{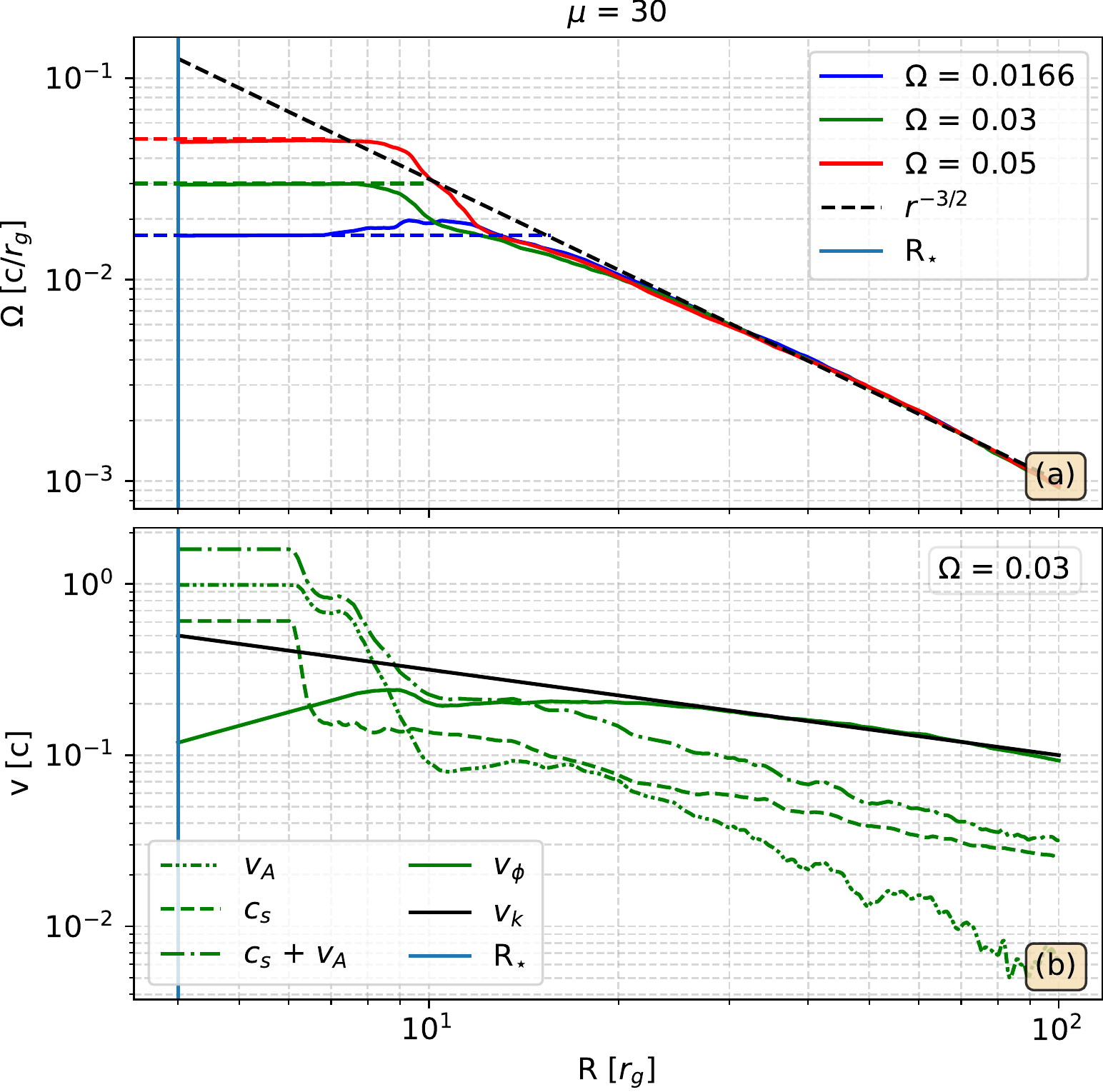}%
    \caption{(a) Solid lines show times averaged omega profiles (t $\in [10\, 000, 30\, 000]r_g/c$ ) in the equatorial plane for the different stellar angular frequencies with dipolar strength of $\mu = 30$ weighted by the metric. The blue, green and red dashed lines show the stellar angular frequencies. The black dashed line shows $r^{-3/2}$ dependence. (b) Different curves represent alfven speed ($v_A$), sound speed ($c_s$), azimuthal velocity ($v_{\phi} = r^2\Omega$) and Kepler velocity in the equatorial plane averaged over t $\in [10\, 000, 30\, 000]r_g/c$.}
    \label{fig:omega}
\end{figure}
To elucidate this point, in Figure \ref{fig:omega} (a) we show the averaged radial dependence of the angular velocity in the equatorial plane for dipolar stellar fields with fixed $\mu=30$ and different stellar rotation parameters $\Omega\in[0.0166,0.03,0.05]$.  All three simulations have a similar magnetospheric radius of $r_m\approx 8r_{g}$. While the rotation at large distances is Keplerian, starting at $\sim 2r_m$, the disk becomes sub-Keplerian.  This is similar to magnetically arrested black hole accretion disks \citep{NarayanIgumenshchev2003,2021MNRAS.502.2023P,BegelmanScepiEtAl2021}, which feature  sub-Keplarian rotation in the inner regions where magnetospheric effects become noticeable. 
The behaviour is further illustrated in the averaged equatorial profiles of sound-speed $c_s = \sqrt{\gamma P / \rho h}$, Alfv\'en velocity $v_A = \sqrt{b^2/(\rho h +b^2)}$ and rotation velocity (Figure \ref{fig:omega} (b)). 
The deviation from the Keplerian profile starts where magnetic and pressure support become comparable to the Kepler speed.  Furthermore, the magnetization in the sub-Keplerian regions is substantial and we obtain plasma-$\beta$ around unity.  
As a consequence of the modified rotation profile of the disk, in all our simulations, the \textit{effective} corotation radius is significantly shifted inwards compared to the expectation.  This is one of the main reasons for the decreased spinup via the magnetic stresses in the accretion column.  

There are several other factors that contribute to the overall spindown of the $\Omega=0.016$ run: first, variations in $r_{\rm m}$ lead to electromagnetic stresses in the column sometimes switching to spin-down, second, the ``disk-connected'' field lines at the column boundaries can impact the disk at radii $r$ with $r_{\rm co}<r<r_{\rm LC}$ and thus exert a strong spin-down visible as steep gradients at the column edge (see e.g. southern column at $\theta\simeq2.8$).  

\begin{figure}
   \centering
    \includegraphics[width=7.2cm]{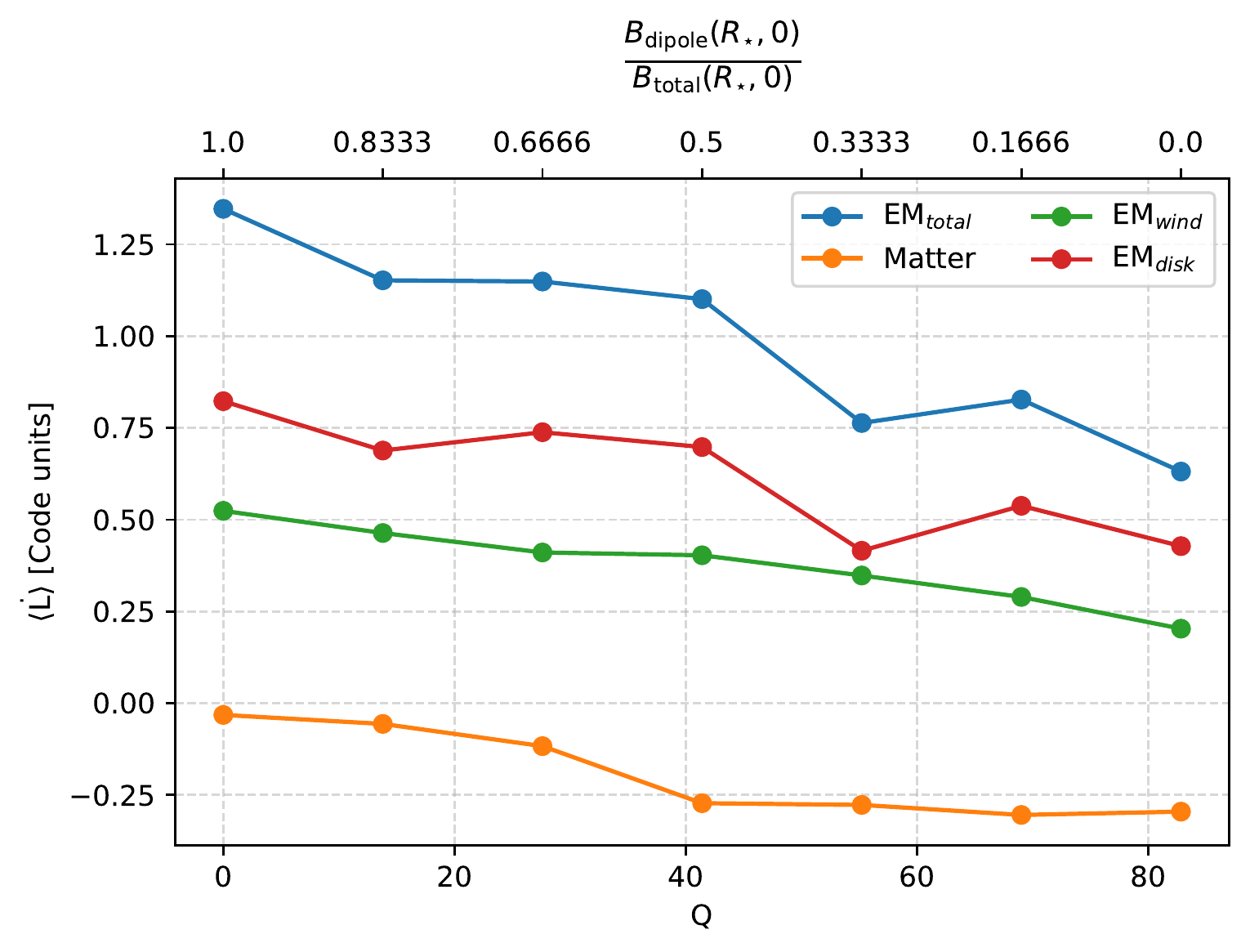}
    \caption{Averaged angular momentum flux for quadrudipoles as a function of increasing quadrupolar contribution. The upper horizontal scale represents the fraction of dipolar field at the stellar surface ($\theta = 0$). All the quantities are averaged over t $\in [10000,30000] r_g/c$.}%
    \label{fig:torquediffQ}
\end{figure}

Figure \ref{fig:torquediffQ} shows the dependence of different components of torque on the quadrupolar strength. For a fair comparison, we increased the quadrupolar contribution and the corresponding dipolar contribution is varied such that effective polar magnetic strength at the stellar surface remains constant. Here Q = 0 represents a pure dipole with $\mu$ = 30 and Q = 82.85 represents a pure quadrupole with same surface field-strength. The total torque decreases with an increase in quadrupolar contribution. 
 
Initially, the matter torque increases as we increase Q (consistent with \citealt{Long2007}) and stabilizes to a constant value for Q $\geq 41.42$. The diminishing increase in matter torque is simultaneous to the development of the accretion belt into a narrow accretion column (cf. Figures \ref{fig:diffQ}, \ref{fig:diffQ-mu}).  This suggests that for large Q, the matter torque is transferred to the disk-connected fields again.  
\begin{figure}
   \centering
    \includegraphics[width=7.2cm]{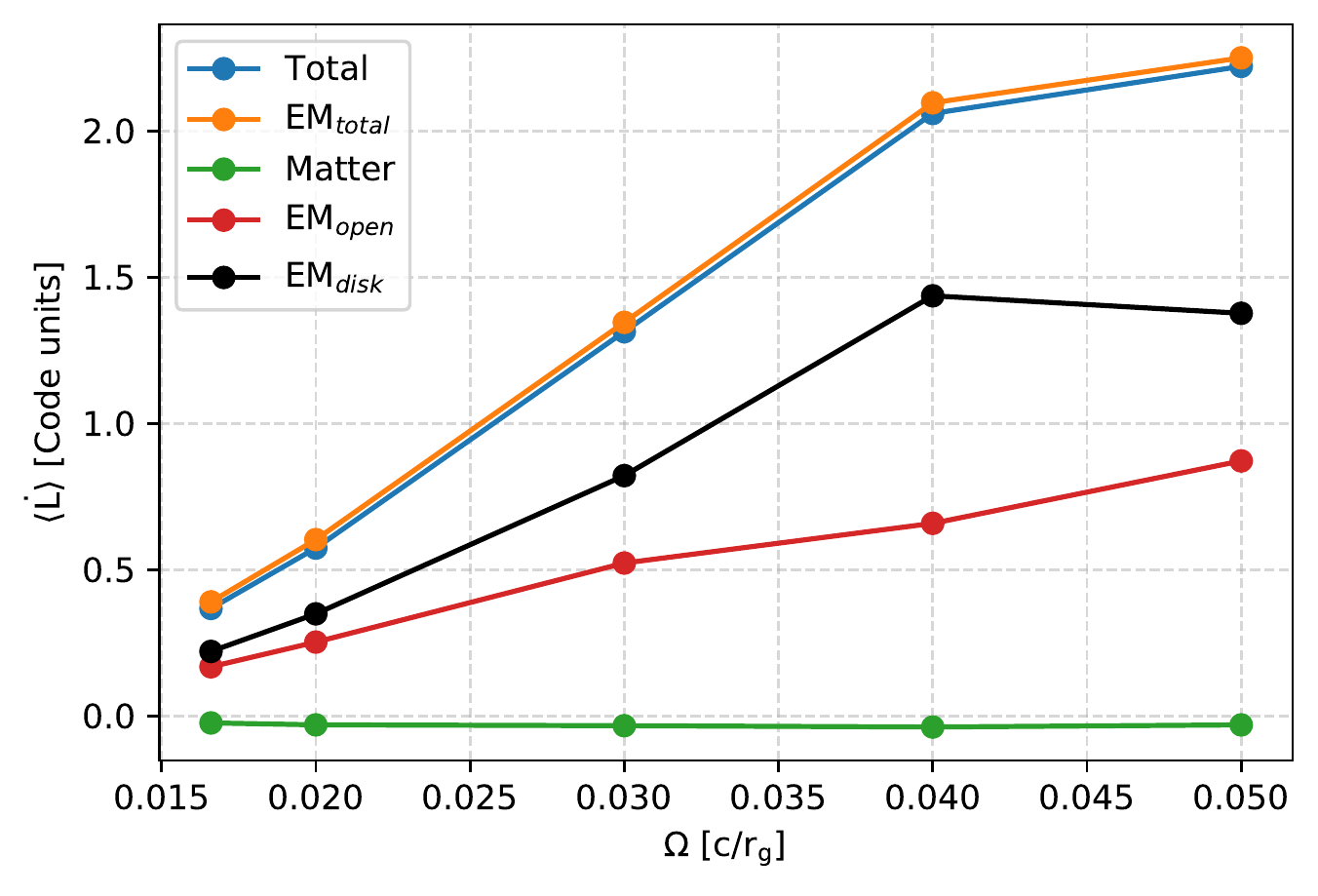}%
    \caption{Angular momentum flux at the stellar surface averaged over $t \in [10000,30000]r_g/c$ as a function of $\Omega$ for dipole  with $\mu$ = 30.} %
    \label{fig:diffwindtorque}
\end{figure}

The net torque in all of the parameter regime and magnetic field geometry explored here is positive, i.e. spinning down the star. Figure \ref{fig:diffwindtorque} illustrates the variation of different components of the torque as a function of $\Omega$ for $\mu$ = 30. As we start increasing the rotation rate, for a fixed magnetic strength ($r_m \approx$ constant), stronger disk-induced spin down results.  However, as the corotation radius approaches the magnetospheric radius for the largest spin value ($\Omega=0.05$), the disk contribution starts to diminish. This behavior is expected as the increase in rotation leads to lesser star-disk connectivity with $\Omega$. Nevertheless, the effective spin-down torque continues to rise with increasing $\Omega$ due to the contribution of the pulsar wind.  
We note that the linear trend in wind torque is consistent with the behavior of the jet power discussed in the next section.  
The matter torque stays negligible and does not vary appreciably with the angular frequency of the star. 
\subsubsection{Stress analysis}
\begin{figure}
   \centering
    \includegraphics[width=8.7cm]{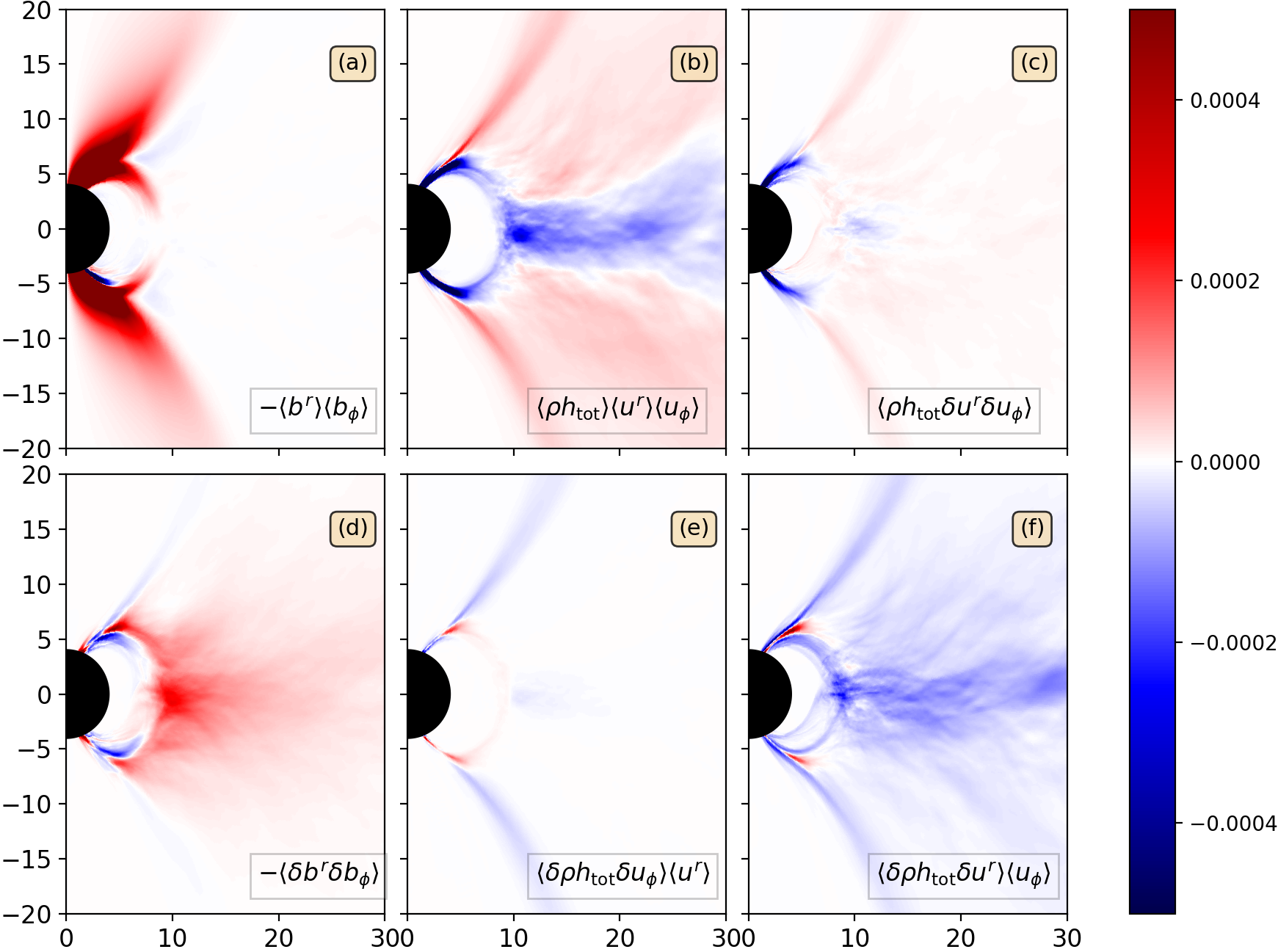}%
     \caption{Different terms in equation \ref{eq:averagedtrphi} for the dipolar fiducial run with $\mu = 30$ with an angular frequency of $\Omega = 0.03$.}
    \label{fig:stress_2d}
\end{figure}

In order to separate out the role of mean and turbulent stresses and investigate the influence of the accretion columns on the disk, we perform a Reynolds decomposition of velocities $u^r$, $u_\phi$, magnetic field components $b^r$, $b_\phi$ as well as $\rho h_{\rm tot}$ whereby $h_{\rm tot} = h + b^2$ is the total specific enthalpy.  The decomposition of an (instantaneous) quantity $X$ is defined as 
\begin{align}
    X = \langle X \rangle + \delta X
\end{align}
where $\langle\cdot\rangle$ denotes the average over time $t\in[10\, 000, 30\, 000]r_g/c$.  This allows to write the angular momentum flux as 
\begin{align}
\rm{T^r_{\phi}} = (\langle \rho h_{tot} \rangle + \delta \rho h_{tot}) (\langle u^r \rangle + \delta u^r) (\langle u_{\phi} \rangle + \delta u_{\phi}) \nonumber \\
- (\langle b^r \rangle + \delta b^r) (\langle b_{\phi} \rangle + \delta b_{\phi})\, . 
\label{eq:averagedtrphi}
\end{align}
Decomposing $\rho h_{\rm tot}$ next to the usual decomposition of the velocity fields that make the Reynolds stress allows us to investigate the role of compressible effects in the angular momentum transport of the thick disk setup.  
The non-vanishing terms in $\langle T^r_{\phi} \rangle$ are shown in Figure \ref{fig:stress_2d}.  
Here, the positive (red) and negative (blue) signs signify angular momentum transport outwards and inwards respectively. Figure \ref{fig:stress_2d} (a) and Figure \ref{fig:stress_2d} (d) show the stresses due to the mean magnetic field and due to its turbulent (Maxwell stress) part. The mean magnetic field stresses resulting from the stellar fields spin down the star and act within the jet and accretion column. The large scale stresses of the column remain fairly localized near $r_{\rm m}\simeq 8 r_g$ and cease entirely at the equatorial plane. The rapid decline of the mean field stress means that there is little evidence for a dynamically significant ``extended magnetosphere'' connecting the star with the disk beyond the corotation radius which would add further magnetic spindown torques \citep{GhoshLamb1979,Zanni2009}. Instead we find that magnetic fields become turbulent leading to dominant Maxwell stresses which provide the largest outwards angular momentum transport throughout the disk.  

Turning to the velocity fields, the mean velocity (Figure \ref{fig:stress_2d} (b)) results in an inward angular momentum transport from accreted material via the equatorial region and the columns (blue regions) as would be expected.  However, we also observe a significant contribution of outwards angular momentum transport via the outflows from the disk. 
Reynolds stresses (Figure \ref{fig:stress_2d} (c)) in the disk mostly transport angular momentum outwards except for the accretion columns where they contribute towards spin-up.  In principle this behavior is expected when the Reynolds stress is modeled as turbulent viscosity. 
Since viscous momentum transport acts always against the gradient of velocity, a decreasing rotation profile within the corotation radius will tend towards spin-up.  
In the simulation shown in Figure \ref{fig:stress_2d} the turbulent contributions in the accretion columns however are minor as the columns are dominated by the mean magnetic field stresses. 

Finally, in panels \ref{fig:stress_2d} (e) and (f), we show correlations of the fluid enthalpy density with the velocity fields.  As the former is dominated by the rest-mass contribution, we are essentially sampling how density fluctuations correlate with the velocity field due to incompressible effects.    
The most striking result is that large densities correlate with large (inflow-) velocities, making a significant contribution to \textit{inward} angular momentum transfer.  We attribute this to \textit{convective motions} whereby dense blobs of gas fall in relatively quickly whereas lighter blobs have the tendency to buoyantly rise.  This behavior can also be observed in the movie of the density field on YOUTUBE \footnote{\url{https://youtu.be/cCl6ZK5vjSM}}.  The fact that thick advection dominated accretion disks are notoriously convectively unstable has been known for some time \citep[][]{NarayanYi1994,QuataertGruzinov2000a,NarayanIgumenshchevEtAl2000}, however strong convective motions have been observed primarily in earlier axisymmetric hydrodynamic studies 
and convection is thought to play little role in the 3D MRI-unstable case \citep{StoneBalbus1996,HawleyBalbusEtAl2001} (though note the more recent work in the magnetically arrested regime by \cite{BegelmanScepiEtAl2021}).  
It will be very interesting to see if the compressible term $\langle \rho h_{\rm tot} \delta u^r\rangle\langle u_\phi\rangle$ remains important in the 3D case.  

\begin{figure}
   \centering
    \includegraphics[width=7cm]{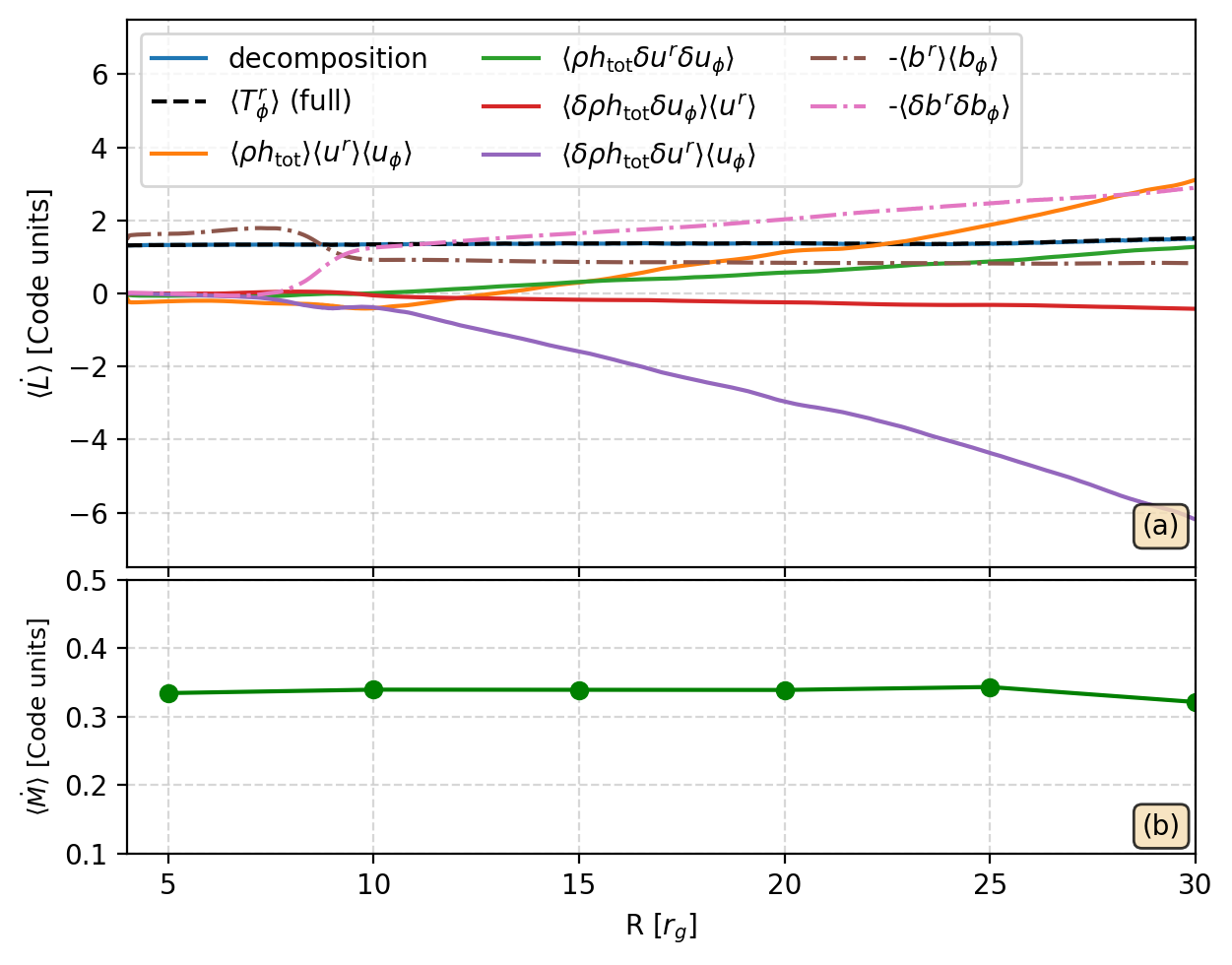}%
    \caption{
     Angular momentum fluxes in the fiducial case with stellar dipole with $\mu = 30$ and $\Omega = 0.03$.
    (a) Radial profiles of different components of the $\rm \langle T^r_{\phi} \rangle$ (Equation \ref{eq:averagedtrphi}) for $t \in [10000,30000]r_g/c$ integrated over $\theta \in [0,\pi]$, (b) Mass accretion rate across shells of the disk showing that a stationary state has been reached in the domain under investigation.}
    \label{fig:stress1d}
\end{figure}

By integrating the individual terms over spherical shells (as in Equation \ref{eq:ldot}), we can assess their contribution to the total torque.  The radial profiles of the fluxes are shown in Figure \ref{fig:stress1d} which also compares $\langle T^r_\phi\rangle$ with and without decomposition (solid blue respectively black dashed curve).  Three points are worth making regarding these two curves:  a) as expected there is net angular momentum transport outwards, b) we obtain excellent agreement between the curves indicating that the decomposition captures all relevant terms and c) the total angular momentum flux is nearly constant throughout the interval $[\rstar,30r_g]$ indicating that these regions have settled into a quasi-stationary state.  The latter point is also evidenced by the constancy of the mass accretion rate shown in the lower panel of Figure \ref{fig:stress1d}.
Large scale magnetic stresses (indicated as brown dot-dashed) provide the dominant contribution within $r_m\simeq8r_g$ where both column- and wind- torque are relevant.  After $\sim10r_g$, a constant spindown contribution from the jet remains.  
As could already be seen in Figure \ref{fig:stress_2d}, turbulent Maxwell stresses govern outwards angular momentum transport throughout the disk, however, we also obtain a large net outwards contribution from to the mean flow $\langle \rho h_{\rm tot}\rangle \langle u^r \rangle\langle u_\phi\rangle$ (solid orange curve).  Hence in the total angular momentum balance, the disk wind at high latitudes dominates over the mass inflow in the equatorial region (cf. Figure \ref{fig:stress_2d} (b)).  Reynolds stresses (solid green) are smaller than Maxwell stresses by a factor $\simeq 3$, which is slightly below the ratio of $4-6$ which is typically found in studies of MRI driven turbulence \citep[e.g.][]{HawleyGammieEtAl1995,PessahEtAl2006,BlackmanPennaEtAl2008,ShiStoneEtAl2016}.  
Next to these well known terms, Figure \ref{fig:stress1d} also shows the contribution of compressible effects which originate in the non-vanishing $\delta(\rho h_{\rm tot})$.  While $\langle \delta \rho h_{\rm tot}\delta u_\phi\rangle \langle u^r\rangle$ remains negligible (solid red line), as seen also in Figure \ref{fig:stress_2d}, the inward transport via $\langle \delta \rho h_{\rm tot}\delta u^r\rangle \langle u_\phi\rangle$ is essential for the total angular momentum balance (solid purple line).  
To our knowledge this term has not been in the focus of previous investigations. It remains to be seen if it also survives in the 3D case as azimuthal shearing motions and magnetic tension from the toroidal field will surely dampen the convective motions responsible for the correlation of density perturbations and inflow velocity seen in our axisymmetric study.

\subsection{Jets}\label{sec:jets}
\begin{figure}
 \centering
 \includegraphics[width=8cm]{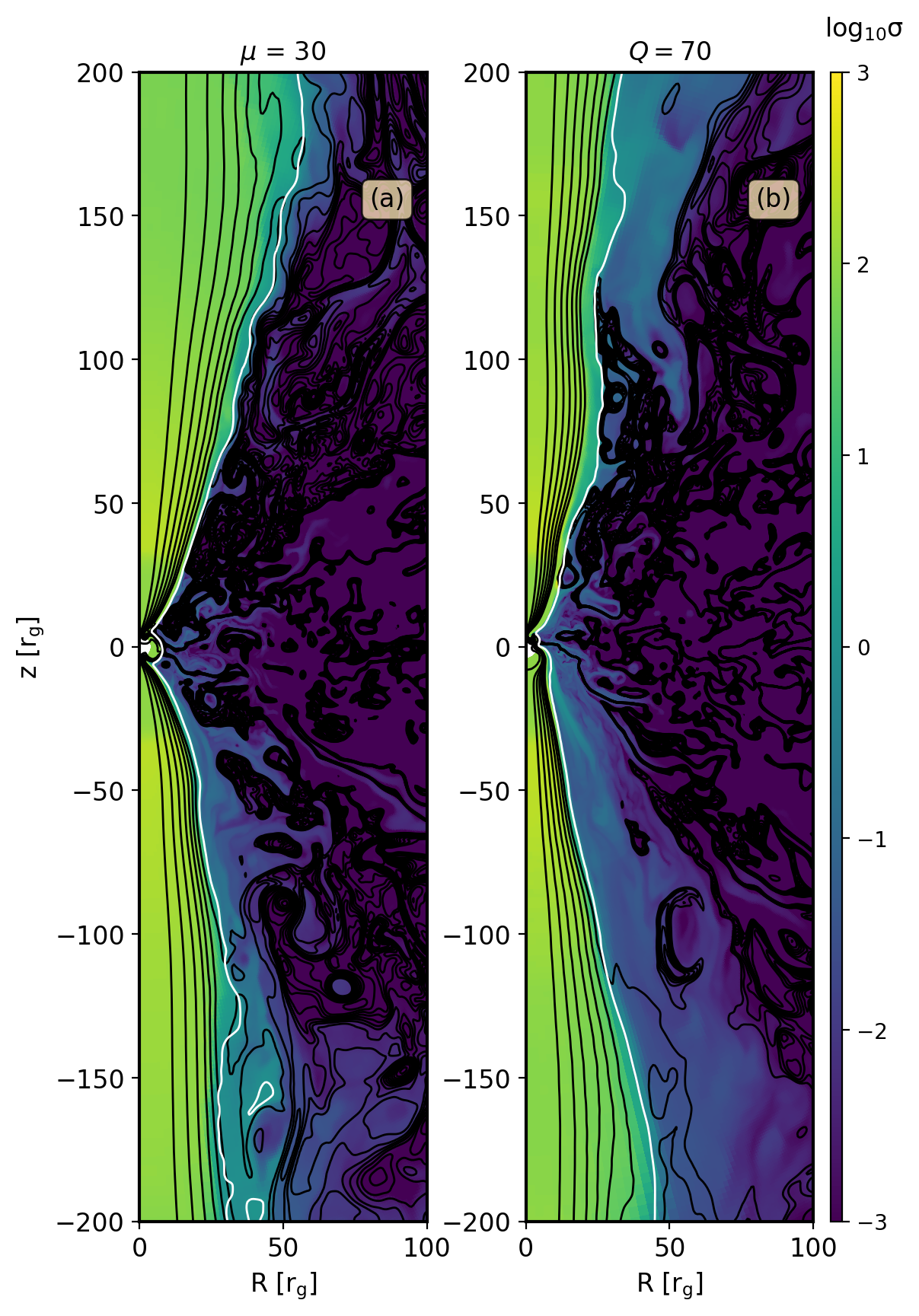}
  \caption{Panel (a) and (b) shows magnetization (log$_{10}\sigma$) for dipole with $\mu = 30$ and quadrupole ($Q = 70$) respectively at $t = 28120 r_g/c$. The black lines represent the contours of $r \sin \theta B^{\phi}$. The jet is defined with the $\sigma = 1$ contour which is highlighted in white.} %
\label{fig:jetstructure}
\end{figure}

\begin{figure}
   \centering
    \includegraphics[width=8.5cm]{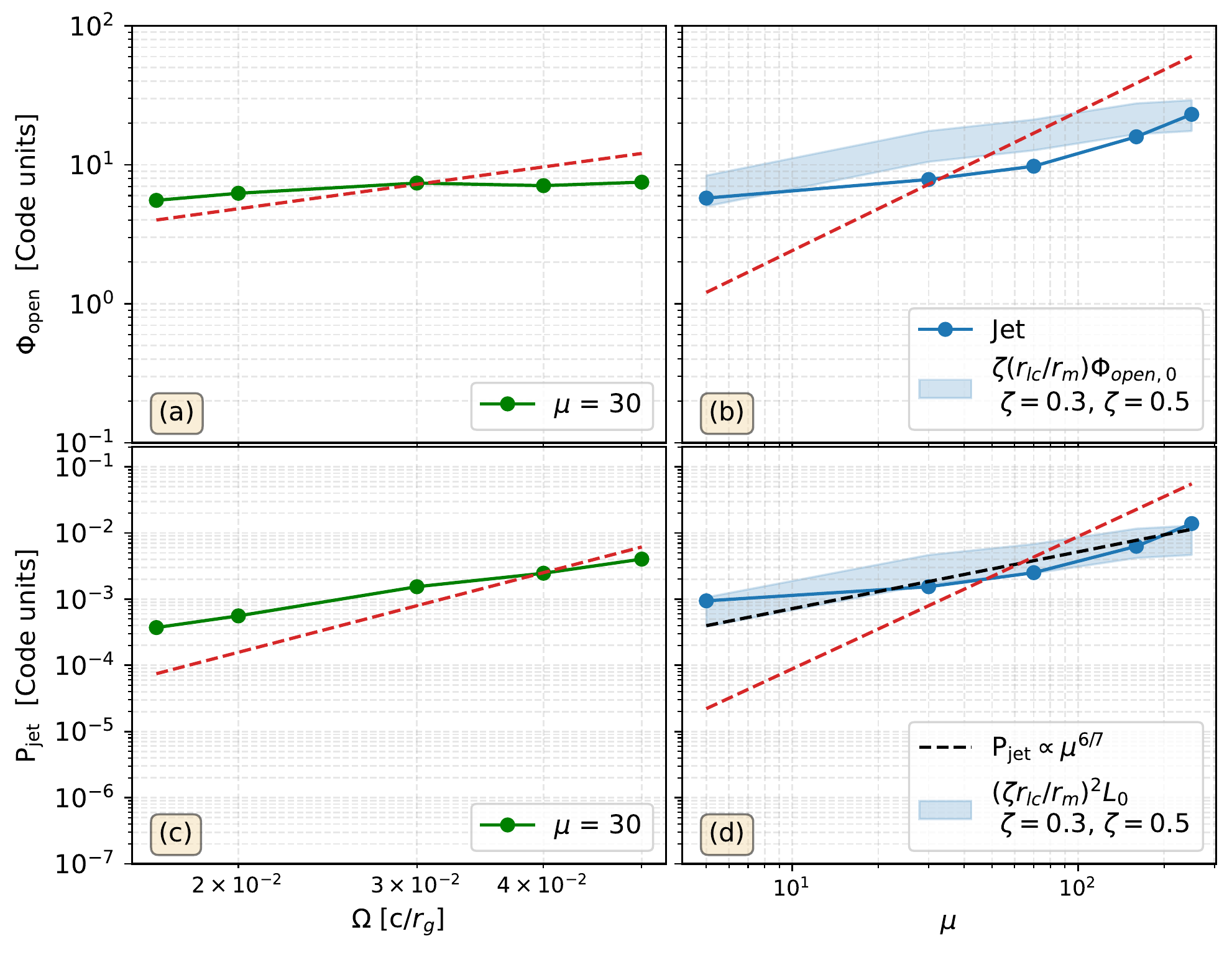}
    \caption{Dipolar open flux (upper panels) and jet power (lower panels) as a function of angular frequency ($\mu$ = 30) and stellar magnetic field ($\Omega = 0.03$). The dashed red lines represent the isolated cases in flat spacetime. The black dashed line in the lower right panel is put arbitrarily to show the expected power-law scaling with surface magnetic strength. The blue shaded regions show the expected jet flux and the jet power from the disk induced flux-opening model.}
    \label{fig:jet-power-di}
\end{figure}
The disk collimates the open polar field lines into a jet. To investigate the dependence of jet properties on parameters, we here define the jet as the magnetically dominated region $\sigma>1$ around the axis. Figure \ref{fig:jetstructure} (a) and (b) illustrates the magnetization profiles ($\sigma$) with contours of $r \sin \theta B^{\phi}$ (black) in the polar regions ($\sigma > 1$) showing the disk induced collimated jet for dipolar and quadrupolar stellar fields respectively. Unlike expected, the highly asymmetric inner magnetosphere for quadrupoles has little impact on the large scale jet structure and is discussed in more details at the end of this section.

\begin{figure}
 \centering
 \includegraphics[width=8.5cm]{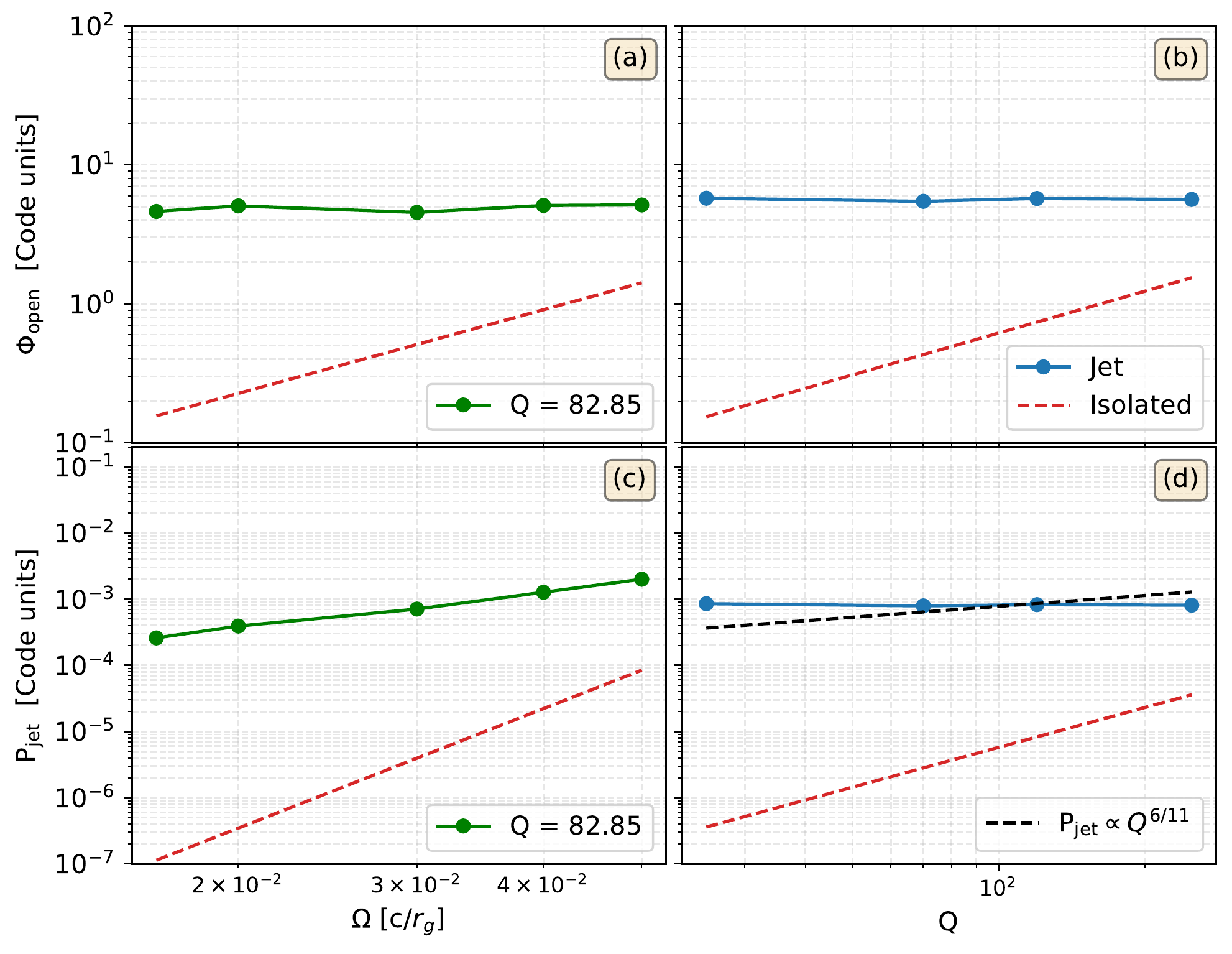}
  \caption{Quadrupolar open flux (upper panels) and jet power (lower panels) as a function of angular frequency (Q = 82.85) and stellar magnetic field ($\Omega = 0.03$). The red dashed lines represent the isolated cases in flat spacetime. The black dashed line in the lower right panel is put arbitrarily to show the expected power-law scaling with surface magnetic strength.} %
\label{fig:jet-power-quad}
\end{figure}

Interaction of the stellar magnetic fields with the disk fields can lead to an enhancement or decrease of the open magnetic flux which comprises the jet.  According to \citetalias{Parfrey2017}, in our dipolar case, where the initial magnetic loop in the disk is anti-parallel to the dipolar stellar field (at the interface between them), a net flux-opening is expected.   This would lead to an enhancement of Poynting-flux compared to the case of the isolated pulsar.  To quantify flux opening and jet power, we place an extraction surface at $r=50r_g$ and integrate Poynting- and magnetic flux for the region $\sigma>1$.  
The variation of open flux ($\Phi_{\rm open}$) and jet power ($P_{\rm jet}$) as a function of the stellar angular frequency and magnetic strength are shown in Figure \ref{fig:jet-power-di} (dipoles) and Figure \ref{fig:jet-power-quad} (quadrupoles). All the quantities are averaged over t $\in$ [10000,30000]$r_g/c$. 
In the far field solution beyond the light-cylinder $r\gg \Omega/c$, one has $E_{\theta}\simeq B_\phi\simeq -(\Omega r\sin{\theta}/c)B_r$ \citep[][]{Vlahakis2004} and hence the Poynting flux can be expressed entirely in terms of open magnetic flux and rotation $\Omega$ \citep[e.g.][]{TchekhovskoyPhilippov2015,Parfrey2016}.  For example, for an isolated rotator in flat spacetime, one finds to good approximation 
\begin{align}
    P_{\rm jet} \approx \frac{2}{3c} \Omega^2 \Phi_{\rm open}^2\, .
\end{align}
Up to a constant coefficient, this scaling (which is analogous to the celebrated \cite{BlandfordZnajek1977} jet-power) is also expected to hold in the case of a collimating jet \citep[][]{TchekhovskoyMcKinney2008}.

Considering first the dipolar case, according to the ``disk induced flux-opening'' model  \citep{MattPudritz2005,Parfrey2016}, the opening for disk-connected fields is determined by the position of the magnetospheric radius to $\Phi_{\rm open} = \zeta r_{\rm lc}/r_m \Phi_{\rm open,0}$ where
$\Phi_{\rm open,0} \mathbf{\approx} 2\pi \mu/r_{lc}$ is the open flux of the corresponding isolated pulsar and so any $\Omega$-dependence cancels out.%
\footnote{The parameter $\zeta<1$ subsumes two effects: for one, as $\Phi_{\rm open,0}$ is obtained from the isolated (dipolar) flux function, it does not take into account compression by the accretion flow.  Compressed field lines crossing $r_m$ in the equatorial plane connect to higher latitudes than the original dipolar field which means that the open flux based on the unmodified dipolar flux function would be overestimated.  A second effect entering into $\zeta$ is the presence of the accretion column (comprised of disk connected field lines) with a width determined by the turbulent magnetic diffusivity of the accretion flow \citep{ParfreySpitkovskyEtAl2017}.}    
For comparison, we show $\Phi_{\rm open,0}$ as red dashed curve in the figures.
We find that for a constant stellar magnetic strength with $r_m < r_{\rm lc}$, the amount of open flux is nearly constant with $\Omega$ giving $\Phi_{\rm open} \propto \Omega^{0.31}$, consistent with the expectation for flux opening at the magnetospheric radius.  
However, as we increase $\Omega$, the expected open flux for the isolated case scaling as $\propto\Omega^1$ overcomes the value of the disk-magnetosphere simulation at $\Omega>0.03$.  This behavior was not reported before (the anti-parallel configurations of \citetalias{Parfrey2017} always show enhanced open flux) and points to a subtle issue of the initialization of the simulations which is discussed further in Section \ref{sec:discussion}.  

Turning to the extracted jet power, Figure \ref{fig:jet-power-di} (c) demonstrates a powerlaw scaling $P_{\rm jet} \propto \Omega^{2.2}$ which is close to the expected slope of $\sim 2.1$ resulting from flux opening.  In particular, the obtained slope is significantly shallower than the isolated case $\propto \Omega^4$.  
The dependence of $\Phi_{\rm open,0}$ and $P_{\rm jet}$ with the stellar field strength is shown in Figure \ref{fig:jet-power-di}(b,d) for fixed $\Omega = 0.03$. 
It can be observed that also here, the open flux and jet power eventually falls behind the isolated pulsar for higher ($\Omega,\mu$). In our case the transition in open flux happens for $\mu > 30$. We interpret the difference in the transition point in $\phi_{\rm open}$ and $P_{\rm jet}$ to be caused by deviations from the monopolar solution in the collimated jet. 
This behavior of the flux opening/closing of the initial flux is corroborated by Figure \ref{fig:diffQ} (c) which shows how the flux contour of the isolated case (pink line) is buried under additionally closed field lines comprising the dead-zone (dark-blue region) out to $r_{\rm m}<r_{\rm lc}$.  Hence the accretion flow has in fact \textit{closed} flux compared to the case of the isolated pulsar with the same parameters. The temporal evolution of open and closed stellar fluxes is exemplified in Appendix \ref{sec:fluxopening}.  

For a more quantitative comparison with the disk induced flux-opening model, we overplot the expected open flux as blue shaded region on the right-hand panels of Figure \ref{fig:jet-power-di}.  
Here we have used the expression for $r_m=0.7 r_{\rm A}$ given by Equation (\ref{eq:rm}) which was found to match well with magnetospheric radii obtained in the simulations (see Section \ref{sec:dipoles}).  This leads to an expectation for $P_{\rm jet} \propto \mu^{6/7} \Omega^2$ which is also shown as a black-dashed line, roughly consistent with the flux opening model for $\zeta \sim 0.3-0.5$.  
Thus given a numerically obtained $\zeta$, the data shows that a reasonable prediction of the jet power is possible once the magnetospheric radius and spin are known.  

Before turning to the data of the quadrupole simulations, let us first provide a simple extension of the disk induced flux-opening model for comparison.  The quadrupole flux follows $\Phi \propto 1/r^2$, hence the total open flux can be written as $\Phi_{\text{open}} = \zeta (r_{\text{LC}}/r_{\text{m}})^2 \Phi_{\text{open,0}}$. Considering the quadrupole scaling of the magnetic pressure, the Alfv\'en radius becomes $r_{\rm A} = ({\rm Q^4/8GM}\dot{\rm M}^2)^{1/11}$ which determines $r_{\rm m} =\xi r_{\rm A}$ ($\xi < 1$) and $\Phi_{\text{open},0} \propto Q\Omega^2/c^2$. This leads to a theoretical jet power estimate of $P_{\text{jet}} \propto Q^{6/11}\Omega^2$ which has the same $\Omega$ dependence but shallower dependence on the stellar field strength compared to the dipole.

Analogue to the dipolar case, Figure \ref{fig:jet-power-quad} shows open flux and jet power for a stellar quadrupole field as a function of $\Omega$ and Q. The total open flux is always above the isolated case suggesting an efficient flux opening in all of the parameter regime considered here. Quantitatively, we obtain $P_{\rm jet} \propto \Omega^{1.8}$ in approximate agreement with the expected powerlaw slope of 2 (whereas the isolated case features a powerlaw slope of $6$).  Just like in the dipole case, the open flux shows essentially no dependence on $\Omega$.

While the nearly flat scaling $\Phi_{\rm open} \propto \Omega^{0.06}$ is expected, we also observe that the open flux (and jet power) is independent of the stellar field strength.  For guidance, the expected slope of $P_{\text{jet}} \propto Q^{6/11}$ is overplotted as a black-dashed line in the bottom right panel of Figure \ref{fig:jet-power-quad}.  
As the expectation also yields a fairly flat slope of $6/11\approx 0.55$, we consider the flatness of the jet power curve with Q noteworthy but likely not yet in strong tension with the theory.  This is especially so when allowing for a similar freedom in the range of the efficiency parameter $\zeta$ as in the dipole cases (blue shaded area in Figure \ref{fig:jet-power-di}). 
Nevertheless, it is important to point out that the disk induced flux opening of the quadrupolar case cannot just be an extension of the dipolar case as the quadrupolar case features both the ``parallel'' (lower hemisphere) and the ``anti-parallel'' configuration (upper hemisphere).  To open up the magnetosphere, stellar field lines first need to \textit{reconnect} to disk field-lines which is favoured in the upper deadzone (see also Figure \ref{fig:diff-disk}). At the same time, the open flux from the lower polar cap is closed due to the presence of the parallel disk field and the equatorial open flux powers the southern jet instead.  

As the inner magnetosphere of the quadrupolar cases is highly asymmetric (c.f. Figure \ref{fig:diffQ}), it is worthwhile considering how this affects the overall symmetry of the jet.  In order to measure the difference in the jet power in the upper and lower hemisphere, we adopt the asymmetry parameter $P_{\rm jet, asymm}$ \citep[][and Appendix \ref{sec:asymm}]{Nathanail2020}. We find small values of the asymmetry $P_{\rm jet, asymm} = 0.08-0.15$ with the upper jet being only $\sim10\%$ stronger than the lower one (see also  Figure \ref{fig:jet-assym-quad}). The disk induced flux opening operates only in the upper hemisphere resulting in slightly higher power in the northern hemisphere.
\subsection{Hotspots}\label{sec:hotspots}

As the channeled matter impacts the stellar surface, hotspots and shocks will form at the base of the accretion columns. 
\footnote{Hotspots and accretion columns in our axisymmetric simulations can be interpreted as hot rings and funnel flows respectively.}
Detailed modeling of the resulting pulsed emission would require us to take into account NS surface models and radiation transport \citep[see for example][]{Poutanen03,Salmi18}. Here, as a first step, we present a simplified treatment of the surface emission following \cite{Romanova2004}. This enables us to explore some of the issues that will be relevant to more sophisticated modeling.

Assuming that the entire matter-kinetic energy flux of the column plasma is emitted as black-body radiation at the stellar surface gives 
\begin{align}
    T_{\rm eff} &= \bigg(\frac{F_{m}}{\sigma}\bigg)^{1/4}
    \end{align}
    where
    \begin{align}
    F_{m} &= {T^{\text{MA}}}^r_t = (\rho_0 + u_g + p_g)u^r u^t 
\end{align}

where $F_{m}$ is the total matter energy flux, $T_{\rm eff}$ and $\sigma$ are the effective black-body temperature and Stefan-Boltzmann constant. Since the average accretion rate (in code units) typically lies within 0.1-0.4 (apart from strong propeller regimes, see Table \ref{tab:runparameters}), we convert the temperature values by assuming that a dimensionless accretion rate of $\dot{M}=0.1$ corresponds to $1\%\dot{M}_{\rm Edd}$ with a radiative efficiency of 0.1.
\begin{figure}
 \centering
\subfloat{{\includegraphics[width=8.7cm]{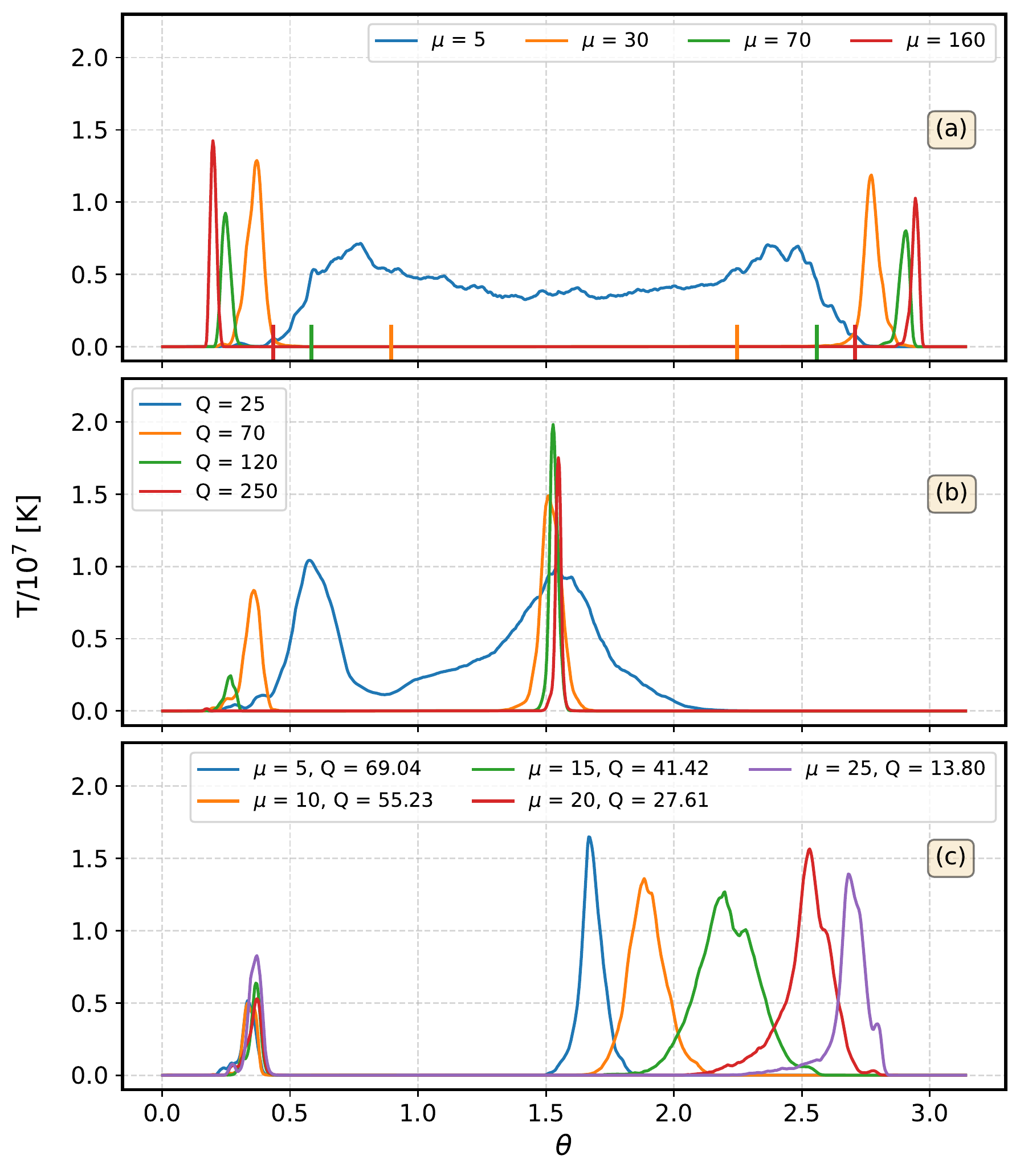} }}%
\caption{Time-averaged hotspot sizes for (a) dipoles, (b) quadrupoles and (c) quadrudipoles for t $\in$ [6000,30000]$r_g/c$. There are clear differences in the degree of channeling for the different magnetic field configurations:  dipoles typically show two clear streams close to the poles; quadrupoles have channeling concentrated in one hemisphere; for quadrudipoles there is at least one strong stream no matter what the configuration but the stream in the other hemisphere is weak. For both quadrupoles and quadrudipoles, the asymmetry between the hemispheres could lead to large differences in pulse visibility depending on observer inclination.}%
\label{fig:hotspots}
\end{figure}

Figure \ref{fig:hotspots} shows the resulting average hotspot temperature as a function of polar angle for different surface magnetic field strengths and geometry.  The average is taken for t $\in [6000,30000] r_g/c$ $\approx$ 200 milliseconds ($\sim$ 100 rotational cycles) for typical $\rstar = 10$km,

For centered dipolar fields, we obtain two accretion columns which are symmetric with respect to the equatorial plane and correspondingly two symmetric hotspots as shown in Figure \ref{fig:hotspots}(a). For sufficiently high magnetic strength ($r_m > \rstar$), a change in $\mu$ leads to a variation in both the location and size of the hotspots. As we increase $\mu$, the magnetic pressure halts the disk at increasingly larger radius ($r_m$ increases) which leads to coupling of accreting matter to stellar fieldlines closer to poles. 
The hotspot is characterized by two angles: $\theta_{\rm min}$ which is given by the last open fieldline and $\theta_{\rm max}$ which is given by the fieldline bordering the deadzone and accretion column. We can qualitatively understand the behavior of these angles as follows: (a) $\theta_{\rm max}$ depends upon the position of the inner disk radius. For an ideal aligned dipole the opening angle can be calculated as $\theta_{\rm max} = {\rm sin}^{-1} (\sqrt{\rstar/r_{\rm m}})$. However, as the disk enters the light-cylinder, it compresses the closed stellar fieldlines towards the neutron star, placing $\theta_{\rm max}$ closer to the poles compared to the ideal dipolar case as discussed in \cite{Zanni2009}. The vertical lines in Figure \ref{fig:hotspots} (a) show the expected $\theta_{\rm max}$ for an ideal dipolar field for different $\mu$. It is roughly twice as far from the pole as the one obtained in our simulations. (b) The position of the last open fieldline which sets $\theta_{\rm min}$ depends on the  connectivity of star-disk fields and on the effective diffusivity of the turbulent disk making it difficult to determine this angle from first principles (see \ref{sec:torque}). However, we note that $\theta_{\rm min}$ is fairly constant with $\mu$ which leads to a decrease in the column thickness with increasing $\mu$.  

Figure \ref{fig:hotspots} (b) represents the hotspot sizes for the quadrupolar cases. Apart from Q = 25, all the quadrupolar field strengths considered here give rise to channeled accretion with two columns, one at the equator and other closer to either of the poles depending upon the relative disk-magnetosphere field orientation (see Figure \ref{fig:diff-disk}). Increasing Q leads to a narrower and stronger equatorial hotspot along with a weaker polar hotspot. Similar to the dipolar case, here too the lower altitude edge of the polar accretion column $\theta_{\rm max}$ is set by the disk truncation radius (in the northern hemisphere in the case shown above).  

As pointed out earlier by \cite{Romanova2004}, we find that the size of the hotspots increases with $\langle \dot{M} \rangle$. One of the common features in both the dipolar and quadrupolar case is the accretion columns becoming thinner with increase in surface stellar magnetic field.

Figure \ref{fig:hotspots}(c) shows the hotspot sizes for quadrudipolar fields with different quadrupolar strengths where the corresponding dipolar field strength is varied such that we end up with the same polar field strength as the pure dipole with $\mu$ = 30. As we increase $\mu$, the equatorial spot becomes broader and shifts towards the lower hemisphere. The shift in the location of the equatorial hotspot towards the lower pole is due to the stellar magnetic field approaching the dipole configuration (Figure \ref{fig:diffQ-mu}). 

\begin{figure*}
\centering
\subfloat{\includegraphics[width=18cm, angle=0]{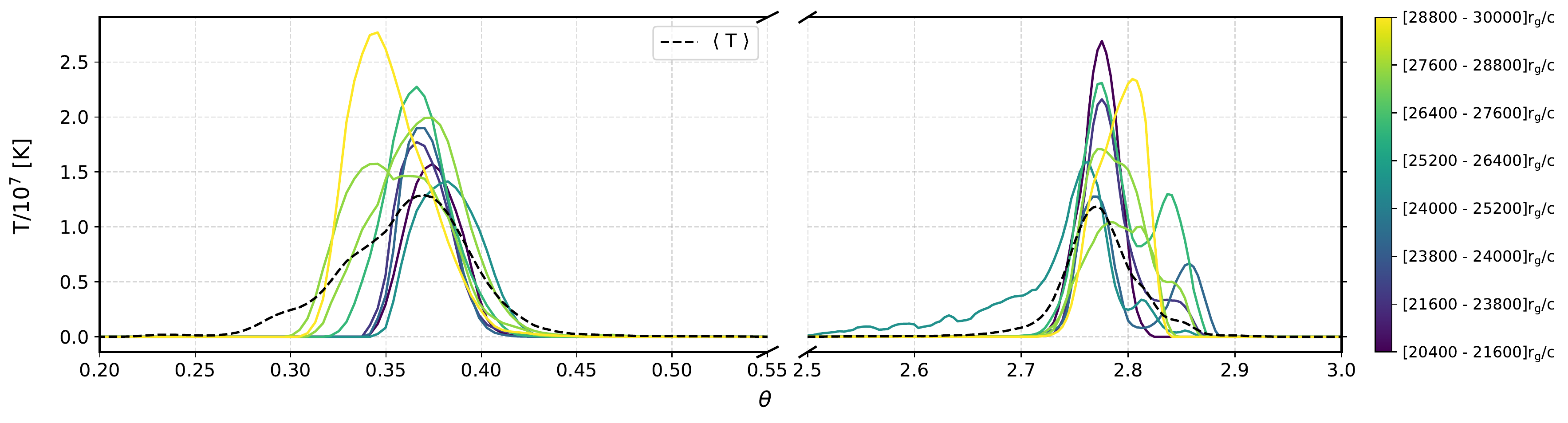}}%
\caption{Temperature profiles at the stellar surface averaged over 1200$r_g/c$ ($\approx$10ms for a typical \rstar = 10 km) for 8 consecutive time windows, for a dipolar field with $\mu$ = 30. The black dashed curve represents the  averaged temperature profile(same as Figure \ref{fig:hotspots} shown for reference). The color gradient shows the temporal progression of the averaged temperature profiles within t $\in$ [20400,30000]$r_g/c \approx$ 80ms.}
\label{fig:mu30-temp}
\end{figure*}

While the previously discussed profiles resulted from a long-term average, in Figure \ref{fig:mu30-temp} we illustrate the temporal variation of the hotspot temperature profiles for the reference case $\mu = 30$. We obtain strong variability of the hotspot sizes and temperatures on timescales as short as $\sim 10\rm ms$.  At the same time, the location of the hotspots given by $\theta_{\rm min}\simeq0.3$ and $\theta_{\rm max}\simeq0.4$ stays approximately constant.  The hotspot sizes in our simulations reflect the size of the accretion column near the stellar surface and the fluctuations in the overall hotspot temperature derive from the strong variations in $\dot{M}$. We also observe instances with multiple peaked structures in the temperature profiles resulting from branching of the accretion column mostly occurring at lower $\dot{M}$.  Such an event is also represented for quadrudipoles in Figure \ref{fig:diffQ-mu}, panels (c,d).  
Whether this variability might be detectable is an interesting question that needs to be explored by more sophisticated models of the emission process that take into account the thermal response time of the surface layers and shock. 

\section{Discussion and Conclusions}\label{sec:discussion}

We have performed a suite of axisymmetric GRMHD simulations to explore accretion onto stars with multipolar surface magnetic field configurations. The initial stellar field strength and angular frequency in our simulations are in the range observed in accreting millisecond pulsars. First, as a reference, we investigated accretion onto stellar dipoles. We recover different accretion states for the dipolar field with anti-parallel star-disk field configuration depending on $\mu$, similar to  \citetalias{Parfrey2017}. The magnetospheric radius for the dipolar case in our simulations for $r_{\rm m} < r_{\rm co}$ is well described by $r_{\rm m} = \xi r_{\rm A}$, with $\xi = 0.7$. For stellar quadrupoles, the presence of parallel and anti-parallel initial star-disk magnetic field configuration in two-hemispheres results in an asymmetric disk-magnetospheric coupling, and in turn, modifies the initial stellar field to resemble a quadrudipolar configuration. For lower field strength, the disk crushes one of the stellar closed zones leading to a slightly asymmetric boundary layer accretion. However, for field strengths typically high enough to balance ram pressure of the disk outside of the star, the accreting materials start to get channeled along the field forming two accretion columns, one at the equator and another closer to the pole. The equatorial hotspot is generally much stronger and less channeled compared to the polar one and this effect increases with Q. In the mixed cases with both dipole and quadrupolar contribution, increasing the dipolar contribution shifts the equatorial column towards the pole, making the inner magnetosphere more symmetric, like the dipolar case.

Considering a star with polar field strength $10^8$ Gauss corresponding to $\mu = 30$, with the scaling choices mentioned in Section \ref{subsec:units}, we get an effective spin-down of $\dot{\nu} \approx 1.8 \times 10^{-15} B_{8}^2 \mu_{30}^{-2}$ Hz/s where $B_{8} = B_{\rm 0,cgs}/10^8\rm G$ and $\mu_{30} = \mu/30$. This is well within the range predicted by the existing torque models, a summary of which can be found in \cite{Parfrey2016}. It is important to mention that in our simulations, the disk connected fieldlines (including the accretion column) always make a significant contribution to the total electromagnetic torque, being either larger or comparable to the wind torque at the stellar surface.  This disk torque component was neglected entirely in the model of \cite{Parfrey2016}.

Significant magnetospheric effects at the inner disk regions leads to sub-Keplarian disks in our simulations. This results in smaller effective corotation radius leading to a net spin down torque for almost all of the parameters explored here. A further investigation into total angular momentum transport shows that although the mean magnetic field stresses dominate angular momentum transport in both accretion columns and the jet, it drops rapidly beyond the magnetospheric radius, suggesting a minimal role of the ``extended disk-magnetosphere'' in the net spin-down observed in our simulations. At the magnetospheric radius, Maxwell stresses dominate immediately leading to an outward angular momentum transport throughout the disk. The mean velocities result in both inward and outward angular momentum transport in the equatorial and wind region. The radial angular momentum flux profile show an increasing contribution of the wind in outward angular momentum transport with radius. The Reynolds stress is smaller than the Maxwell stress by a factor of $\sim 3$ and leads to an overall outward angular momentum transport in the disk. We find that a strong (inward) contribution to the total angular momentum transport in the disk comes from the correlation between density and radial velocity fluctuations which can be explained by a scenario where denser plasma has a stronger tendency to rapid inward motion than lighter buoyant gas -- as in convection. However, whether these correlated density-radial velocity fluctuations survive in absence of axisymmetry needs further study by means of 3D simulations.

The spin-down torque decreases due to the presence of non-dipolar fields at the stellar surface. We find that for sufficiently low stellar frequency, the net torque at the star varies significantly depending on initial field geometry. At a fixed stellar frequency, more quadrupolar contribution decreases the spin-down EM torque and increases the matter torque (which is always spin-up) simultaneously, leading to a net decrease in the total spin-down torque at the star. 
The effect of different stellar field geometry on the torque was previously explored by \citet{Long2007,Long2008} where an increase in quadrupolar contribution instead lead to less spin-up torque. The difference in behaviour can be understood as follows: (a) the higher $\dot{M}$ considered in \cite{Long2007} results in a spin-up EM torque compared to the spin-down contribution in our case; (b) their thin non-magnetised disk retains the initial quadrupolar field which leads to only one accretion column through the equator, whereas in our simulations, in case of a pure quadrupole, the disk field reconnects with the initial stellar geometry to form two accretion columns, one through the equator and other closer to the pole. This increases the star-disk connectivity in our case, leading to transfer of angular momentum to the field at an effective magnetospheric radius that results in a dominant EM spin-down torque. This EM spin-down torque is however smaller than in the corresponding dipole case.  Thus when the quadrupolar contribution is increased (while keeping the polar field strength fixed), we obtain a decrease in spin-down EM torque and an increase in the spin-up matter torque, leading to a net decrease in total spin-down torque at the star.                       

As a possible caveat of our analysis we should mention that the separation criterion for the wind zone at the star (see Section \ref{sec:torque}) is dependent on the $\sigma$ threshold ($\sigma_{\rm cut} = 1$) where increasing $\sigma_{\rm cut}$ can lead to a slight decrease in the wind contribution. 
We checked that the open flux corresponding to our measured wind torque agrees with the jet flux (defined with $\sigma =1$ contour and extracted at $r=50 r_{\rm g}$) to within $\approx 10\%$ showing that what we denote as ``wind'' will indeed reach large distances. The wind zone contours on the torque $\theta$ profiles at the stellar surface (Figure \ref{fig:mu30-om0p0166-om0p03}), along with the conservation of open-flux, supports our threshold choice for the disk- and wind-torque separation.    

We find that the presence of the accretion flow significantly modifies the open magnetic flux, torque and jet power compared to the case of the isolated pulsar.  For the configuration studied here, where disk field and stellar field are anti-parallel (at their intersection), field-lines can become connected to the accretion disk via magnetic reconnection.  The differential footpoint rotation then leads to ballooning of the field lines, which quickly rip apart to generate new open magnetic flux.  The latter process was described in a number of works, for example  \citet{Uzdensky2002a,Uzdensky2002b,LyndenBell2003,Parfrey2016,Ferreira2000}.  Our results for the dipolar stellar field are qualitatively consistent with the ``disk induced flux opening model'' of \cite{Parfrey2016}, that is, with reference to the isolated spindown power $L_0\simeq\mu^2\Omega^4/c^3$ \citep[][]{Gruzinov2005}, the enhanced power due to disk induced open flux is determined by $P_{\rm jet}=(\zeta r_{\rm LC}/r_{\rm m})^2 L_0$ which results in a parameter dependence $\propto\mu^{6/7}\Omega^{2}$.  
The best fitting slope for the spin in our simulations is $P_{\rm jet}\propto \Omega^{1.8}$, very near this expectation, and the scaling with $\mu$ is recovered when allowing for a range in $\zeta$. 
We note that our obtained efficiency parameter $0.3<\zeta<0.5$ is lower than in the similar study by \citetalias{Parfrey2017} who quote $0.5<\zeta<0.75$.  More importantly, we obtain a different scaling of the jet power with spin frequency: \citetalias{Parfrey2017} report the steeper dependence $P_{\rm jet}\propto \Omega^{3.6}$ which is closer to the scaling of the isolated dipole $\propto\Omega^4$.  
Further, we observe for our high spin and magnetic field simulations that magnetic flux is in fact \textit{closed} compared to the isolated pulsar case.  This is theoretically possible since the accretion flow compresses the closed zone to within the initial $r_{\rm lc}$ in the isolated regime.  In principle, this behaviour is expected only for ``parallel'' configurations, which points to a subtle dependence on the initial conditions discussed further below.  

Flux opening in the quadrupole configuration proceeds differently to the dipolar case: by simply extending the analysis of the disk induced flux opening model to quadrupoles, we obtain an intrinsically weaker dependence of the open polar field strength than in the dipole case ($\propto \mu^{6/7}$ vs. $\propto Q^{6/11}$).  
More importantly, quadrupole stellar fields give rise to both ``anti-parallel'' and ``parallel'' regions, depending on the hemisphere under consideration.  It is hence far from obvious whether and how flux opening proceeds in the quadrupolar case. Our numerical experiments demonstrate that flux opening is also efficient in case of quadrupoles and turns out to be independent of the strength of the stellar field, in contrast to the case of stellar dipoles.  

While the upper hemisphere jet emanates from field lines rooted in the polar region (``isolated'' ones plus newly opened up ones), the lower hemisphere jet is composed of field-lines rooted in the equatorial region.  These shield the accretion flow from reaching the star and thus disallow the formation of an accretion column in the lower hemisphere. 
This geometry of the open field lines gives rise to a slight asymmetry ($\sim 10\%$) in jet power from quadrupole stellar fields.   

Formation of asymmetric and even one-sided jets from complex stellar fields was previously reported by \citet{Lovelace2010}.  In their study, a thin viscoresistive accretion disk interacted with  quadru-dipolar stellar fields with varying contributions.  Most notably, their pure dipolar configurations displayed one-sidedness and a cyclic flip-flop behavior. None of our simulations exhibit such time-dependence, which seems to imply that the mechanism is limited to thin-disks only.  When the disk scale height is large, we expect that the compression of the stellar field is more homogeneous, which avoids the creation of energetically favored upward- or downward- accretion paths.  

If we scale our fiducial case ($\mu=30,\ \Omega=0.03$) to polar field strengths of $10^9\rm G$, we obtain a jet power of $\simeq 1.7\times 10^{35} \rm erg s^{-1}$, sufficient to explain the inferred radio jet power from the neutron star X-ray binaries Cir X-1 and Sco X-1 of $>10^{35} \rm erg s^{-1}$ \citep{FenderWuEtAl2004}. Importantly, we find that the case with quadrupolar stellar fields of the same polar field strength (Q$\simeq83$) emits a comparable amount of power.  In both cases, the required accretion luminosity stays below the Eddington limit with $L_{\rm Acc}/L_{\rm Edd}\simeq5\%$.  Allowing our model to accrete at the Eddington limit (as expected for Sco X-1, \citealt{BradshawFomalontEtAl1999}) and using the scaling of the jet power $\propto\Omega^2$, the power limit $>10^{35}\rm erg s^{-1}$ can be met for spin frequencies $<7\rm ms$.  

Recently, radio jets have also been detected from strongly magnetized ($>10^{12}\rm G$) neutron stars with periods above several seconds \citep{Eijnden2018,Eijnden2021}.  However, with merely $\sim10^{28}\rm erg\, s^{-1}$ their typical $6\rm GHz $ radio power is substantially lower than the inferred jet power of Sco X-1.  In the case of Swift J0243.6+6124 (hereafter SW J0243), the spin period is close to $10\rm s$ and a magnetic field strength of approximately $10^{13}\rm G$ was inferred from modeling of the spinup torque \citep{DoroshenkoEtAl2018Fixed}.  The X-ray luminosity places the source at or above the Eddington limit \citep{Eijnden2018}.  
Assuming that the scaling for the jet power in the dipolar case obtained in Section \ref{sec:jets} also holds at the Eddington limit, we can write in convenient units
\begin{align}
    P_{\rm jet} \simeq 9.7\times 10^{32} \left(\frac{B_{0,\rm cgs}}{10^{12}\rm G}\right)^{6/7} \left(\frac{P}{1\rm s}\right)^{-2}\rm erg\, s^{-1}
\end{align}
which yields $\simeq 7\times 10^{31}\rm erg\, s^{-1}$ for the parameters of SW J0243.  Thus the radio power can easily be explained by our model and requires only a small radiative efficiency $\sim 10^{-4}$.  

Since the dependence of the jet power on the surface field is much weaker for stellar quadrupoles (while they share the same $P^{-2}$ period dependence), it is interesting to ask whether such field topologies can be ruled out by underpredicting the radio power of objects like SW J0243.  This is however not the case: even if we assume the jet power is independent of the surface field strength, scaled to SW J0243, the jet power from the quadrupole stellar field case still yields $\simeq 8.4\times 10^{28}\rm erg\, s^{-1}$.  If we adopt the Q$^{6/11}$ dependence instead, the power falls just an order of magnitude short of the dipolar scaling, thus still allowing enough room for the radio emission.   

One striking difference of our results compared to \citetalias{Parfrey2017} is that although we consider the anti-parallel case, our simulations do not uniformly open up stellar magnetic flux, but the cases with $\mu>50$ lead to more closed flux compared to the case of the isolated pulsar (see e.g. panel (b) of Figure \ref{fig:jet-power-di}).  This can also be seen in Figure \ref{fig:diffmudipole} where the flux contour of the isolated case is buried under additional closed field lines. 
The reason for this extra closed flux can be understood when inspecting the initial transient phase of the simulations: as the stellar rotation is switched on, an Alfv\'en wave sweeps through the domain which pushes the magnetic flux initially located between the light-cylinder and the torus towards the torus edge. These trapped fieldlines being parallel to the stellar closed zone field represent excess flux that can close through the Y-point when the deadzone is pushed by the accretion flow within the light-cylinder.  As the strength of the stellar field is increased, also the contribution of the ``excess flux'' increases.  The initial excess flux is negligible in the case of stellar quadrupole fields as these have vanishing meridional fields at the equator. Correspondingly, we always recover significant flux opening in the quadrupolar cases. 
In the setup of \citetalias{Parfrey2017}, we expect that this effect is diminished as the initial stellar dipole field is modified to ``loop around'' the torus.  

This somewhat subtle issue points out the dependence of the simulation on the initial conditions.  It has already been demonstrated by \citetalias{Parfrey2017} that parallel and anti-parallel configurations yield vastly different results, with the parallel cases leading to a suppression of the stellar jet.  
The variant described here indicates once more that the amount of open flux and whether or not a jet will be formed in the accreting system depends on the initial conditions (or accretion history) of the object.  
In the field of black hole accretion simulations, the dependence on the initial magnetic field configuration in the accretion flow has been known for some time \citep[see][and references therein]{Komissarov2021} and the generation and transport of large scale magnetic fields is an important open question.  Unsurprisingly, it seems that simulations of accreting neutron stars are no different.  

How accretion is channeled onto the star's surface determines whether or not it will be detectable as an X-ray pulsar. Our simulations show that where hotspots would form, and their size and temperature distribution, depends strongly on magnetic field configuration.  For quadrupolar and quadrudipolar configurations, there are pronounced differences in the flows impacting the two different hemispheres: observer inclination would then be even more important in determining whether the detected emission is pulsed.  We also find high levels of variability in the channeled accretion flows, something that may contribute to the high frequency variability and timing noise in X-ray emission from accreting neutron stars \citep[see for example][]{Hartman2008,Bachetti2010,PatrunoandWatts2012,Mendez2021}. There are also implications for future attempts to do pulse profile modeling to measure mass and radius of accreting neutron stars. Pulse profile modeling uses physically-motivated models for the generation of the pulsed emission. Pulse profiles (detected counts, resolved in rotational phase and energy) are built up over many rotational cycles, in order to obtain sufficient numbers of photons for the analysis.  If the underlying pulse generation process is highly variable, we will need to verify whether this affects our ability to recover model parameters successfully.  However a full assessment of the degree of variability in pulse generation will require 3D simulations and better models of the emission process. 

This axisymmetric GRMHD study of accreting neutron stars has shown that magnetic field geometries more complex than a dipole can have major effects on channeling of accretion flows. Our next step will be to extend this study to 3D, where it will be interesting to see whether the same features that we have observed in 2D persist in the higher dimensional problem.  We also need to consider disk thickness in more detail.  The thick disk simulations presented here should be appropriate for the low-luminosity hard-state or the super-Eddington regime, as argued in \citetalias{Parfrey2017} and \cite{ParfreySpitkovskyEtAl2017}. At the low accretion rate end this would cover some of the accretion rates observed for the accretion powered millisecond pulsars, which typically accrete at less than 10\% of the Eddington rate \citep{PatrunoandWatts2012}.  However many accreting neutron stars, including those with thermonuclear bursts but without detected accretion-powered pulsations, have intermediate accretion rates \citep{Galloway2020}.  Modeling this accretion rate regime will require the investigation of thin accretion disks, which we also plan to do in a follow up study. 

\section*{Acknowledgements}

P.D. and O.P. acknowledge funding from  the Virtual Institute for Accretion (VIA) within NOVA (Nederlandse Onderzoeksschool voor Astronomie) Network 3 ``Astrophysics in extreme conditions''. Simulations have been carried out in part on the HELIOS cluster of the Anton Pannekoek Institute for Astronomy and on the Dutch national e-infrastructure with the support of SURF Cooperative.  A.L.W. acknowledges support from ERC Consolidator Grant No.~865768 AEONS.  We thank Gibwa Musoke and Tuomo Salmi for comments. We also thank Bart Ripperda and Nathalie Degenaar for interesting discussions. 

\section*{Data Availability}

The data underlying this article are available in Zenodo at \url{https://doi.org/10.5281/zenodo.5674918}

\typeout{}
\bibliographystyle{mnras}
\bibliography{References}

\begin{thebibliography}{}
\makeatletter
\relax
\def\mn@urlcharsother{\let\do\@makeother \do\$\do\&\do\#\do\^\do\_\do\%\do\~}
\def\mn@doi{\begingroup\mn@urlcharsother \@ifnextchar [ {\mn@doi@}
  {\mn@doi@[]}}
\def\mn@doi@[#1]#2{\def\@tempa{#1}\ifx\@tempa\@empty \href
  {http://dx.doi.org/#2} {doi:#2}\else \href {http://dx.doi.org/#2} {#1}\fi
  \endgroup}
\def\mn@eprint#1#2{\mn@eprint@#1:#2::\@nil}
\def\mn@eprint@arXiv#1{\href {http://arxiv.org/abs/#1} {{\tt arXiv:#1}}}
\def\mn@eprint@dblp#1{\href {http://dblp.uni-trier.de/rec/bibtex/#1.xml}
  {dblp:#1}}
\def\mn@eprint@#1:#2:#3:#4\@nil{\def\@tempa {#1}\def\@tempb {#2}\def\@tempc
  {#3}\ifx \@tempc \@empty \let \@tempc \@tempb \let \@tempb \@tempa \fi \ifx
  \@tempb \@empty \def\@tempb {arXiv}\fi \@ifundefined
  {mn@eprint@\@tempb}{\@tempb:\@tempc}{\expandafter \expandafter \csname
  mn@eprint@\@tempb\endcsname \expandafter{\@tempc}}}

\bibitem[\protect\citeauthoryear{{Alpar}, {Cheng}, {Ruderman}  \&
  {Shaham}}{{Alpar} et~al.}{1982}]{Alpar1982}
{Alpar} M.~A.,  {Cheng} A.~F.,  {Ruderman} M.~A.,   {Shaham} J.,  1982, \mn@doi
  [\nat] {10.1038/300728a0}, \href
  {https://ui.adsabs.harvard.edu/abs/1982Natur.300..728A} {300, 728}

\bibitem[\protect\citeauthoryear{{Andersson}, {Glampedakis}, {Haskell}  \&
  {Watts}}{{Andersson} et~al.}{2005}]{Andersson2005}
{Andersson} N.,  {Glampedakis} K.,  {Haskell} B.,   {Watts} A.~L.,  2005,
  \mn@doi [\mnras] {10.1111/j.1365-2966.2005.09167.x}, \href
  {https://ui.adsabs.harvard.edu/abs/2005MNRAS.361.1153A} {361, 1153}

\bibitem[\protect\citeauthoryear{{Ardeljan}, {Bisnovatyi-Kogan}  \&
  {Moiseenko}}{{Ardeljan} et~al.}{2005}]{Ardeljan2005}
{Ardeljan} N.~V.,  {Bisnovatyi-Kogan} G.~S.,   {Moiseenko} S.~G.,  2005,
  \mn@doi [\mnras] {10.1111/j.1365-2966.2005.08888.x}, \href
  {https://ui.adsabs.harvard.edu/abs/2005MNRAS.359..333A} {359, 333}

\bibitem[\protect\citeauthoryear{{Bachetti}, {Romanova}, {Kulkarni}, {Burderi}
  \& {di Salvo}}{{Bachetti} et~al.}{2010}]{Bachetti2010}
{Bachetti} M.,  {Romanova} M.~M.,  {Kulkarni} A.,  {Burderi} L.,   {di Salvo}
  T.,  2010, \mn@doi [\mnras] {10.1111/j.1365-2966.2010.16203.x}, \href
  {https://ui.adsabs.harvard.edu/abs/2010MNRAS.403.1193B} {403, 1193}

\bibitem[\protect\citeauthoryear{{Barnard} \& {Arons}}{{Barnard} \&
  {Arons}}{1982}]{Barnard1982}
{Barnard} J.~J.,  {Arons} J.,  1982, \mn@doi [\apj] {10.1086/159784}, \href
  {https://ui.adsabs.harvard.edu/abs/1982ApJ...254..713B} {254, 713}

\bibitem[\protect\citeauthoryear{Begelman, Scepi  \& Dexter}{Begelman
  et~al.}{2021}]{BegelmanScepiEtAl2021}
Begelman M.~C.,  Scepi N.,   Dexter J.,  2021, What Really Makes an Accretion
  Disc {{MAD}}, \url {https://ui.adsabs.harvard.edu/abs/2021arXiv211102439B}

\bibitem[\protect\citeauthoryear{{Beskin}}{{Beskin}}{2018}]{Beskin2018}
{Beskin} V.~S.,  2018, \mn@doi [Physics Uspekhi] {10.3367/UFNe.2017.10.038216},
  \href {https://ui.adsabs.harvard.edu/abs/2018PhyU...61..353B} {61, 353}

\bibitem[\protect\citeauthoryear{{Bessolaz}, {Zanni}, {Ferreira}, {Keppens}  \&
  {Bouvier}}{{Bessolaz} et~al.}{2008}]{Bessolaz2008}
{Bessolaz} N.,  {Zanni} C.,  {Ferreira} J.,  {Keppens} R.,   {Bouvier} J.,
  2008, \mn@doi [\aap] {10.1051/0004-6361:20078328}, \href
  {https://ui.adsabs.harvard.edu/abs/2008A&A...478..155B} {478, 155}

\bibitem[\protect\citeauthoryear{{Bhattacharya} \& {van den
  Heuvel}}{{Bhattacharya} \& {van den Heuvel}}{1991}]{Bhattacharya1991}
{Bhattacharya} D.,  {van den Heuvel} E.~P.~J.,  1991, \mn@doi [\physrep]
  {10.1016/0370-1573(91)90064-S}, \href
  {https://ui.adsabs.harvard.edu/abs/1991PhR...203....1B} {203, 1}

\bibitem[\protect\citeauthoryear{{Bilous} et~al.,}{{Bilous}
  et~al.}{2019}]{Bilous2019}
{Bilous} A.~V.,  et~al., 2019, \mn@doi [\apjl] {10.3847/2041-8213/ab53e7},
  \href {https://ui.adsabs.harvard.edu/abs/2019ApJ...887L..23B} {887, L23}

\bibitem[\protect\citeauthoryear{Blackman, Penna  \& Varni{\`e}re}{Blackman
  et~al.}{2008}]{BlackmanPennaEtAl2008}
Blackman E.~G.,  Penna R.~F.,   Varni{\`e}re P.,  2008, \mn@doi [New Astronomy]
  {10.1016/j.newast.2007.10.004}, 13, 244

\bibitem[\protect\citeauthoryear{Blandford \& Znajek}{Blandford \&
  Znajek}{1977}]{BlandfordZnajek1977}
Blandford R.~D.,  Znajek R.~L.,  1977, Monthly Notices of the Royal
  Astronomical Society, 179, 433

\bibitem[\protect\citeauthoryear{{Bogovalov}}{{Bogovalov}}{1997}]{Bogovalov1997fixed}
{Bogovalov} S.~V.,  1997, \aap, \href
  {https://ui.adsabs.harvard.edu/abs/1997A&A...323..634B} {323, 634}

\bibitem[\protect\citeauthoryear{{Bradshaw}, {Fomalont}  \&
  {Geldzahler}}{{Bradshaw} et~al.}{1999}]{BradshawFomalontEtAl1999}
{Bradshaw} C.~F.,  {Fomalont} E.~B.,   {Geldzahler} B.~J.,  1999, \mn@doi
  [\apjl] {10.1086/311889}, \href
  {https://ui.adsabs.harvard.edu/abs/1999ApJ...512L.121B} {512, L121}

\bibitem[\protect\citeauthoryear{{Di Salvo} \& {Sanna}}{{Di Salvo} \&
  {Sanna}}{2020}]{DiSalvo20}
{Di Salvo} T.,  {Sanna} A.,  2020, arXiv e-prints, \href
  {https://ui.adsabs.harvard.edu/abs/2020arXiv201009005D} {p. arXiv:2010.09005}

\bibitem[\protect\citeauthoryear{{Donati} et~al.,}{{Donati}
  et~al.}{2007}]{Donati2007}
{Donati} J.~F.,  et~al., 2007, \mn@doi [\mnras]
  {10.1111/j.1365-2966.2007.12194.x}, \href
  {https://ui.adsabs.harvard.edu/abs/2007MNRAS.380.1297D} {380, 1297}

\bibitem[\protect\citeauthoryear{{Donati} et~al.,}{{Donati}
  et~al.}{2008}]{Donati2008}
{Donati} J.~F.,  et~al., 2008, \mn@doi [\mnras]
  {10.1111/j.1365-2966.2008.13111.x}, \href
  {https://ui.adsabs.harvard.edu/abs/2008MNRAS.386.1234D} {386, 1234}

\bibitem[\protect\citeauthoryear{{Doroshenko}, {Tsygankov}  \&
  {Santangelo}}{{Doroshenko} et~al.}{2018}]{DoroshenkoEtAl2018Fixed}
{Doroshenko} V.,  {Tsygankov} S.,   {Santangelo} A.,  2018, \mn@doi [\aap]
  {10.1051/0004-6361/201732208}, \href
  {https://ui.adsabs.harvard.edu/abs/2018A&A...613A..19D} {613, A19}

\bibitem[\protect\citeauthoryear{{Elsner} \& {Lamb}}{{Elsner} \&
  {Lamb}}{1977}]{Elsner1977}
{Elsner} R.~F.,  {Lamb} F.~K.,  1977, \mn@doi [\apj] {10.1086/155427}, \href
  {https://ui.adsabs.harvard.edu/abs/1977ApJ...215..897E} {215, 897}

\bibitem[\protect\citeauthoryear{{Fender}, {Wu}, {Johnston}, {Tzioumis},
  {Jonker}, {Spencer}  \& {van der Klis}}{{Fender}
  et~al.}{2004}]{FenderWuEtAl2004}
{Fender} R.,  {Wu} K.,  {Johnston} H.,  {Tzioumis} T.,  {Jonker} P.,  {Spencer}
  R.,   {van der Klis} M.,  2004, \mn@doi [\nat] {10.1038/nature02137}, \href
  {https://ui.adsabs.harvard.edu/abs/2004Natur.427..222F} {427, 222}

\bibitem[\protect\citeauthoryear{{Fendt}}{{Fendt}}{2009}]{Fendt2009conf}
{Fendt} C.,  2009, in Protostellar Jets in Context. pp 131--136 (\mn@eprint
  {arXiv} {0811.3416}), \mn@doi{10.1007/978-3-642-00576-3\_16}

\bibitem[\protect\citeauthoryear{{Ferreira}, {Pelletier}  \& {Appl}}{{Ferreira}
  et~al.}{2000}]{Ferreira2000}
{Ferreira} J.,  {Pelletier} G.,   {Appl} S.,  2000, \mn@doi [\mnras]
  {10.1046/j.1365-8711.2000.03215.x}, \href
  {https://ui.adsabs.harvard.edu/abs/2000MNRAS.312..387F} {312, 387}

\bibitem[\protect\citeauthoryear{{Fishbone} \& {Moncrief}}{{Fishbone} \&
  {Moncrief}}{1976}]{Fishbone1976}
{Fishbone} L.~G.,  {Moncrief} V.,  1976, \mn@doi [\apj] {10.1086/154565}, \href
  {https://ui.adsabs.harvard.edu/abs/1976ApJ...207..962F} {207, 962}

\bibitem[\protect\citeauthoryear{{Galloway} \& {Keek}}{{Galloway} \&
  {Keek}}{2021}]{GallowayKeek2021}
{Galloway} D.~K.,  {Keek} L.,  2021, in {Belloni} T.~M.,  {M{\'e}ndez} M.,
  {Zhang} C.,  eds,  Astrophysics and Space Science Library Vol. 461,
  Astrophysics and Space Science Library. pp 209--262 (\mn@eprint {arXiv}
  {1712.06227}), \mn@doi{10.1007/978-3-662-62110-3\_5}

\bibitem[\protect\citeauthoryear{{Galloway} et~al.,}{{Galloway}
  et~al.}{2020}]{Galloway2020}
{Galloway} D.~K.,  et~al., 2020, \mn@doi [\apjs] {10.3847/1538-4365/ab9f2e},
  \href {https://ui.adsabs.harvard.edu/abs/2020ApJS..249...32G} {249, 32}

\bibitem[\protect\citeauthoryear{Ghosh \& Lamb}{Ghosh \&
  Lamb}{1979}]{GhoshLamb1979}
Ghosh P.,  Lamb F.~K.,  1979, \mn@doi [The Astrophysical Journal]
  {10.1086/157498}, 234, 296

\bibitem[\protect\citeauthoryear{{G{\"o}{\v{g}}{\"u}{\c{s}}}, {Alpar}  \&
  {Gilfanov}}{{G{\"o}{\v{g}}{\"u}{\c{s}}} et~al.}{2007}]{Gogus2007}
{G{\"o}{\v{g}}{\"u}{\c{s}}} E.,  {Alpar} M.~A.,   {Gilfanov} M.,  2007, \mn@doi
  [\apj] {10.1086/512028}, \href
  {https://ui.adsabs.harvard.edu/abs/2007ApJ...659..580G} {659, 580}

\bibitem[\protect\citeauthoryear{{Gralla}, {Lupsasca}  \& {Philippov}}{{Gralla}
  et~al.}{2017}]{Gralla2017}
{Gralla} S.~E.,  {Lupsasca} A.,   {Philippov} A.,  2017, \mn@doi [\apj]
  {10.3847/1538-4357/aa978d}, \href
  {https://ui.adsabs.harvard.edu/abs/2017ApJ...851..137G} {851, 137}

\bibitem[\protect\citeauthoryear{Gruzinov}{Gruzinov}{1999}]{Gruzinov1999}
Gruzinov A.,  1999, arXiv e-prints, pp astro--ph/9902288

\bibitem[\protect\citeauthoryear{Gruzinov}{Gruzinov}{2005}]{Gruzinov2005}
Gruzinov A.,  2005, \mn@doi [\prl] {10.1103/PhysRevLett.94.021101}, 94, 021101

\bibitem[\protect\citeauthoryear{{Harding} \& {Muslimov}}{{Harding} \&
  {Muslimov}}{2011}]{HardingMuslimov2011fixed}
{Harding} A.~K.,  {Muslimov} A.~G.,  2011, \mn@doi [\apjl]
  {10.1088/2041-8205/726/1/L10}, \href
  {https://ui.adsabs.harvard.edu/abs/2011ApJ...726L..10H} {726, L10}

\bibitem[\protect\citeauthoryear{{Hartman} et~al.,}{{Hartman}
  et~al.}{2008}]{Hartman2008}
{Hartman} J.~M.,  et~al., 2008, \mn@doi [\apj] {10.1086/527461}, \href
  {https://ui.adsabs.harvard.edu/abs/2008ApJ...675.1468H} {675, 1468}

\bibitem[\protect\citeauthoryear{Hawley, Gammie  \& Balbus}{Hawley
  et~al.}{1995}]{HawleyGammieEtAl1995}
Hawley J.~F.,  Gammie C.~F.,   Balbus S.~A.,  1995, \mn@doi [The Astrophysical
  Journal] {10.1086/175311}, 440, 742

\bibitem[\protect\citeauthoryear{Hawley, Balbus  \& Stone}{Hawley
  et~al.}{2001}]{HawleyBalbusEtAl2001}
Hawley J.~F.,  Balbus S.~A.,   Stone J.~M.,  2001, \mn@doi [The Astrophysical
  Journal] {10.1086/320931}, 554, L49

\bibitem[\protect\citeauthoryear{{Igoshev}, {Popov}  \& {Hollerbach}}{{Igoshev}
  et~al.}{2021}]{Igoshev2021}
{Igoshev} A.~P.,  {Popov} S.~B.,   {Hollerbach} R.,  2021, arXiv e-prints,
  \href {https://ui.adsabs.harvard.edu/abs/2021arXiv210905584I} {p.
  arXiv:2109.05584}

\bibitem[\protect\citeauthoryear{{Jones}}{{Jones}}{1980}]{Jones1980}
{Jones} P.~B.,  1980, \mn@doi [\mnras] {10.1093/mnras/192.4.847}, \href
  {https://ui.adsabs.harvard.edu/abs/1980MNRAS.192..847J} {192, 847}

\bibitem[\protect\citeauthoryear{{Keek} et~al.,}{{Keek}
  et~al.}{2018}]{KeekArzoumanianEtAl2018}
{Keek} L.,  et~al., 2018, \mn@doi [\apjl] {10.3847/2041-8213/aab904}, \href
  {https://ui.adsabs.harvard.edu/abs/2018ApJ...856L..37K} {856, L37}

\bibitem[\protect\citeauthoryear{Komissarov}{Komissarov}{2002}]{Komissarov2002}
Komissarov S.~S.,  2002, \mn@doi [Monthly Notices of the Royal Astronomical
  Society] {10.1046/j.1365-8711.2002.05313.x}, 336, 759

\bibitem[\protect\citeauthoryear{{Komissarov}}{{Komissarov}}{2006}]{komissarov2006}
{Komissarov} S.~S.,  2006, \mn@doi [\mnras] {10.1111/j.1365-2966.2005.09932.x},
  \href {https://ui.adsabs.harvard.edu/abs/2006MNRAS.367...19K} {367, 19}

\bibitem[\protect\citeauthoryear{{Komissarov} \& {Porth}}{{Komissarov} \&
  {Porth}}{2021}]{Komissarov2021}
{Komissarov} S.,  {Porth} O.,  2021, \mn@doi [\nar]
  {10.1016/j.newar.2021.101610}, \href
  {https://ui.adsabs.harvard.edu/abs/2021NewAR..9201610K} {92, 101610}

\bibitem[\protect\citeauthoryear{{Konar}}{{Konar}}{2017}]{Konar2017}
{Konar} S.,  2017, \mn@doi [Journal of Astrophysics and Astronomy]
  {10.1007/s12036-017-9467-4}, \href
  {https://ui.adsabs.harvard.edu/abs/2017JApA...38...47K} {38, 47}

\bibitem[\protect\citeauthoryear{{Kulkarni} \& {Romanova}}{{Kulkarni} \&
  {Romanova}}{2005}]{Kulkarni2005}
{Kulkarni} A.~K.,  {Romanova} M.~M.,  2005, \mn@doi [\apj] {10.1086/444489},
  \href {https://ui.adsabs.harvard.edu/abs/2005ApJ...633..349K} {633, 349}

\bibitem[\protect\citeauthoryear{{Kulkarni} \& {Romanova}}{{Kulkarni} \&
  {Romanova}}{2008}]{Kulkarni2008}
{Kulkarni} A.~K.,  {Romanova} M.~M.,  2008, \mn@doi [\mnras]
  {10.1111/j.1365-2966.2008.13094.x}, \href
  {https://ui.adsabs.harvard.edu/abs/2008MNRAS.386..673K} {386, 673}

\bibitem[\protect\citeauthoryear{Kylafis, Tr{\"u}mper  \& Loudas}{Kylafis
  et~al.}{2021}]{KylafisTrumperEtAl2021a}
Kylafis N.~D.,  Tr{\"u}mper J.~E.,   Loudas N.~A.,  2021, \mn@doi [Astronomy
  and Astrophysics] {10.1051/0004-6361/202039361}, 655, A39

\bibitem[\protect\citeauthoryear{{Lamb}, {Boutloukos}, {Van Wassenhove},
  {Chamberlain}, {Lo}, {Clare}, {Yu}  \& {Miller}}{{Lamb}
  et~al.}{2009}]{Lamb2009}
{Lamb} F.~K.,  {Boutloukos} S.,  {Van Wassenhove} S.,  {Chamberlain} R.~T.,
  {Lo} K.~H.,  {Clare} A.,  {Yu} W.,   {Miller} M.~C.,  2009, \mn@doi [\apj]
  {10.1088/0004-637X/706/1/417}, \href
  {https://ui.adsabs.harvard.edu/abs/2009ApJ...706..417L} {706, 417}

\bibitem[\protect\citeauthoryear{{Long}, {Romanova}  \& {Lovelace}}{{Long}
  et~al.}{2007}]{Long2007}
{Long} M.,  {Romanova} M.~M.,   {Lovelace} R.~V.~E.,  2007, \mn@doi [\mnras]
  {10.1111/j.1365-2966.2006.11192.x}, \href
  {https://ui.adsabs.harvard.edu/abs/2007MNRAS.374..436L} {374, 436}

\bibitem[\protect\citeauthoryear{{Long}, {Romanova}  \& {Lovelace}}{{Long}
  et~al.}{2008}]{Long2008}
{Long} M.,  {Romanova} M.~M.,   {Lovelace} R.~V.~E.,  2008, \mn@doi [\mnras]
  {10.1111/j.1365-2966.2008.13124.x}, \href
  {https://ui.adsabs.harvard.edu/abs/2008MNRAS.386.1274L} {386, 1274}

\bibitem[\protect\citeauthoryear{{Lovelace}, {Romanova}, {Ustyugova}  \&
  {Koldoba}}{{Lovelace} et~al.}{2010}]{Lovelace2010}
{Lovelace} R.~V.~E.,  {Romanova} M.~M.,  {Ustyugova} G.~V.,   {Koldoba} A.~V.,
  2010, \mn@doi [\mnras] {10.1111/j.1365-2966.2010.17284.x}, \href
  {https://ui.adsabs.harvard.edu/abs/2010MNRAS.408.2083L} {408, 2083}

\bibitem[\protect\citeauthoryear{{Lynden-Bell}}{{Lynden-Bell}}{2003}]{LyndenBell2003}
{Lynden-Bell} D.,  2003, \mn@doi [\mnras] {10.1046/j.1365-8711.2003.06506.x},
  \href {https://ui.adsabs.harvard.edu/abs/2003MNRAS.341.1360L} {341, 1360}

\bibitem[\protect\citeauthoryear{Matt \& Pudritz}{Matt \&
  Pudritz}{2005a}]{MattPudritz2005a}
Matt S.,  Pudritz R.~E.,  2005a, \mn@doi [Monthly Notices of the Royal
  Astronomical Society] {10.1111/j.1365-2966.2004.08431.x}, 356, 167

\bibitem[\protect\citeauthoryear{Matt \& Pudritz}{Matt \&
  Pudritz}{2005b}]{MattPudritz2005}
Matt S.,  Pudritz R.~E.,  2005b, \mn@doi [The Astrophysical Journal]
  {10.1086/498066}, 632, L135

\bibitem[\protect\citeauthoryear{McKinney}{McKinney}{2006}]{McKinney2006c}
McKinney J.~C.,  2006, \mn@doi [Monthly Notices of the Royal Astronomical
  Society] {10.1111/j.1745-3933.2006.00150.x}, 368, L30

\bibitem[\protect\citeauthoryear{{M{\'e}ndez} \& {Belloni}}{{M{\'e}ndez} \&
  {Belloni}}{2021}]{Mendez2021}
{M{\'e}ndez} M.,  {Belloni} T.~M.,  2021, in {Belloni} T.~M.,  {M{\'e}ndez} M.,
    {Zhang} C.,  eds,  Astrophysics and Space Science Library Vol. 461,
  Astrophysics and Space Science Library. pp 263--331 (\mn@eprint {arXiv}
  {2010.08291}), \mn@doi{10.1007/978-3-662-62110-3\_6}

\bibitem[\protect\citeauthoryear{{Miller} et~al.,}{{Miller}
  et~al.}{2019}]{Miller2019}
{Miller} M.~C.,  et~al., 2019, \mn@doi [\apjl] {10.3847/2041-8213/ab50c5},
  \href {https://ui.adsabs.harvard.edu/abs/2019ApJ...887L..24M} {887, L24}

\bibitem[\protect\citeauthoryear{{Miller} et~al.,}{{Miller}
  et~al.}{2021}]{Miller2021}
{Miller} M.~C.,  et~al., 2021, \mn@doi [\apjl] {10.3847/2041-8213/ac089b},
  \href {https://ui.adsabs.harvard.edu/abs/2021ApJ...918L..28M} {918, L28}

\bibitem[\protect\citeauthoryear{Narayan \& Yi}{Narayan \&
  Yi}{1994}]{NarayanYi1994}
Narayan R.,  Yi I.,  1994, \mn@doi [The Astrophysical Journal]
  {10.1086/187381}, 428, L13

\bibitem[\protect\citeauthoryear{Narayan, Igumenshchev  \& Abramowicz}{Narayan
  et~al.}{2000}]{NarayanIgumenshchevEtAl2000}
Narayan R.,  Igumenshchev I.~V.,   Abramowicz M.~A.,  2000, \mn@doi [The
  Astrophysical Journal] {10.1086/309268}, 539, 798

\bibitem[\protect\citeauthoryear{{Narayan}, {Igumenshchev}  \&
  {Abramowicz}}{{Narayan} et~al.}{2003}]{NarayanIgumenshchev2003}
{Narayan} R.,  {Igumenshchev} I.~V.,   {Abramowicz} M.~A.,  2003, \mn@doi
  [\pasj] {10.1093/pasj/55.6.L69}, \href
  {https://ui.adsabs.harvard.edu/abs/2003PASJ...55L..69N} {55, L69}

\bibitem[\protect\citeauthoryear{{Nathanail}, {Fromm}, {Porth}, {Olivares},
  {Younsi}, {Mizuno}  \& {Rezzolla}}{{Nathanail} et~al.}{2020}]{Nathanail2020}
{Nathanail} A.,  {Fromm} C.~M.,  {Porth} O.,  {Olivares} H.,  {Younsi} Z.,
  {Mizuno} Y.,   {Rezzolla} L.,  2020, \mn@doi [\mnras]
  {10.1093/mnras/staa1165}, \href
  {https://ui.adsabs.harvard.edu/abs/2020MNRAS.495.1549N} {495, 1549}

\bibitem[\protect\citeauthoryear{{Obergaulinger} \& {Aloy}}{{Obergaulinger} \&
  {Aloy}}{2017}]{Obergaulinger2017}
{Obergaulinger} M.,  {Aloy} M.~{\'A}.,  2017, in Journal of Physics Conference
  Series. p. 012043 (\mn@eprint {arXiv} {1711.09975}),
  \mn@doi{10.1088/1742-6596/932/1/012043}

\bibitem[\protect\citeauthoryear{{Olivares}, {Porth}, {Davelaar}, {Most},
  {Fromm}, {Mizuno}, {Younsi}  \& {Rezzolla}}{{Olivares}
  et~al.}{2019}]{Olivares2019}
{Olivares} H.,  {Porth} O.,  {Davelaar} J.,  {Most} E.~R.,  {Fromm} C.~M.,
  {Mizuno} Y.,  {Younsi} Z.,   {Rezzolla} L.,  2019, \mn@doi [\aap]
  {10.1051/0004-6361/201935559}, \href
  {https://ui.adsabs.harvard.edu/abs/2019A&A...629A..61O} {629, A61}

\bibitem[\protect\citeauthoryear{{{\"O}zel}}{{{\"O}zel}}{2009}]{Ozel2009}
{{\"O}zel} F.,  2009, \mn@doi [\apj] {10.1088/0004-637X/691/2/1678}, \href
  {https://ui.adsabs.harvard.edu/abs/2009ApJ...691.1678O} {691, 1678}

\bibitem[\protect\citeauthoryear{{Parfrey} \& {Tchekhovskoy}}{{Parfrey} \&
  {Tchekhovskoy}}{2017}]{Parfrey2017}
{Parfrey} K.,  {Tchekhovskoy} A.,  2017, \mn@doi [\apjl]
  {10.3847/2041-8213/aa9c85}, \href
  {https://ui.adsabs.harvard.edu/abs/2017ApJ...851L..34P} {851, L34}

\bibitem[\protect\citeauthoryear{{Parfrey}, {Spitkovsky}  \&
  {Beloborodov}}{{Parfrey} et~al.}{2016}]{Parfrey2016}
{Parfrey} K.,  {Spitkovsky} A.,   {Beloborodov} A.~M.,  2016, \mn@doi [\apj]
  {10.3847/0004-637X/822/1/33}, \href
  {https://ui.adsabs.harvard.edu/abs/2016ApJ...822...33P} {822, 33}

\bibitem[\protect\citeauthoryear{{Parfrey}, {Spitkovsky}  \&
  {Beloborodov}}{{Parfrey} et~al.}{2017}]{ParfreySpitkovskyEtAl2017}
{Parfrey} K.,  {Spitkovsky} A.,   {Beloborodov} A.~M.,  2017, \mn@doi [\mnras]
  {10.1093/mnras/stx950}, \href
  {https://ui.adsabs.harvard.edu/abs/2017MNRAS.469.3656P} {469, 3656}

\bibitem[\protect\citeauthoryear{{Patruno} \& {Watts}}{{Patruno} \&
  {Watts}}{2021}]{PatrunoandWatts2012}
{Patruno} A.,  {Watts} A.~L.,  2021, \mn@doi [Astrophysics and Space Science
  Library] {10.1007/978-3-662-62110-3\_4}, \href
  {https://ui.adsabs.harvard.edu/abs/2021ASSL..461..143P} {461, 143}

\bibitem[\protect\citeauthoryear{Pessah, Chan  \& Psaltis}{Pessah
  et~al.}{2006}]{PessahEtAl2006}
Pessah M.~E.,  Chan C.-K.,   Psaltis D.,  2006, \mn@doi [Monthly Notices of the
  Royal Astronomical Society] {10.1111/j.1365-2966.2006.10824.x}, 372, 183

\bibitem[\protect\citeauthoryear{{P{\'e}tri}}{{P{\'e}tri}}{2021}]{Petri2021}
{P{\'e}tri} J.,  2021, \mn@doi [\mnras] {10.1093/mnras/staa3909}, \href
  {https://ui.adsabs.harvard.edu/abs/2021MNRAS.501.4479P} {501, 4479}

\bibitem[\protect\citeauthoryear{{Porth} \& {Fendt}}{{Porth} \&
  {Fendt}}{2010}]{Porth2010}
{Porth} O.,  {Fendt} C.,  2010, \mn@doi [\apj] {10.1088/0004-637X/709/2/1100},
  \href {https://ui.adsabs.harvard.edu/abs/2010ApJ...709.1100P} {709, 1100}

\bibitem[\protect\citeauthoryear{{Porth}, {Olivares}, {Mizuno}, {Younsi},
  {Rezzolla}, {Moscibrodzka}, {Falcke}  \& {Kramer}}{{Porth}
  et~al.}{2017}]{Porth2017}
{Porth} O.,  {Olivares} H.,  {Mizuno} Y.,  {Younsi} Z.,  {Rezzolla} L.,
  {Moscibrodzka} M.,  {Falcke} H.,   {Kramer} M.,  2017, \mn@doi [Computational
  Astrophysics and Cosmology] {10.1186/s40668-017-0020-2}, \href
  {https://ui.adsabs.harvard.edu/abs/2017ComAC...4....1P} {4, 1}

\bibitem[\protect\citeauthoryear{{Porth}, {Chatterjee}, {Narayan}, {Gammie},
  {Mizuno}, {Anninos}  \& {Event Horizon Telescope Collaboration}}{{Porth}
  et~al.}{2019}]{Porth2019}
{Porth} O.,  {Chatterjee} K.,  {Narayan} R.,  {Gammie} C.~F.,  {Mizuno} Y.,
  {Anninos} P.,   {Event Horizon Telescope Collaboration} 2019, \mn@doi [\apjs]
  {10.3847/1538-4365/ab29fd}, \href
  {https://ui.adsabs.harvard.edu/abs/2019ApJS..243...26P} {243, 26}

\bibitem[\protect\citeauthoryear{Porth, Mizuno, Younsi  \& Fromm}{Porth
  et~al.}{2021}]{2021MNRAS.502.2023P}
Porth O.,  Mizuno Y.,  Younsi Z.,   Fromm C.~M.,  2021, \mn@doi [Monthly
  Notices of the Royal Astronomical Society] {10.1093/mnras/stab163}, 502, 2023

\bibitem[\protect\citeauthoryear{{Poutanen} \& {Gierli{\'n}ski}}{{Poutanen} \&
  {Gierli{\'n}ski}}{2003}]{Poutanen03}
{Poutanen} J.,  {Gierli{\'n}ski} M.,  2003, \mn@doi [\mnras]
  {10.1046/j.1365-8711.2003.06773.x}, \href
  {https://ui.adsabs.harvard.edu/abs/2003MNRAS.343.1301P} {343, 1301}

\bibitem[\protect\citeauthoryear{{Psaltis} \& {Chakrabarty}}{{Psaltis} \&
  {Chakrabarty}}{1999}]{Psaltis1999b}
{Psaltis} D.,  {Chakrabarty} D.,  1999, \mn@doi [\apj] {10.1086/307525}, \href
  {https://ui.adsabs.harvard.edu/abs/1999ApJ...521..332P} {521, 332}

\bibitem[\protect\citeauthoryear{Quataert \& Gruzinov}{Quataert \&
  Gruzinov}{2000}]{QuataertGruzinov2000a}
Quataert E.,  Gruzinov A.,  2000, \mn@doi [The Astrophysical Journal]
  {10.1086/309267}, 539, 809

\bibitem[\protect\citeauthoryear{{Radhakrishnan} \&
  {Srinivasan}}{{Radhakrishnan} \& {Srinivasan}}{1982}]{Radhakrishnan1982}
{Radhakrishnan} V.,  {Srinivasan} G.,  1982, Current Science, \href
  {https://ui.adsabs.harvard.edu/abs/1982CSci...51.1096R} {51, 1096}

\bibitem[\protect\citeauthoryear{{Riley} et~al.,}{{Riley}
  et~al.}{2019}]{Riley2019}
{Riley} T.~E.,  et~al., 2019, \mn@doi [\apjl] {10.3847/2041-8213/ab481c}, \href
  {https://ui.adsabs.harvard.edu/abs/2019ApJ...887L..21R} {887, L21}

\bibitem[\protect\citeauthoryear{{Riley} et~al.,}{{Riley}
  et~al.}{2021}]{Riley2021}
{Riley} T.~E.,  et~al., 2021, \mn@doi [\apjl] {10.3847/2041-8213/ac0a81}, \href
  {https://ui.adsabs.harvard.edu/abs/2021ApJ...918L..27R} {918, L27}

\bibitem[\protect\citeauthoryear{{Romanova}, {Ustyugova}, {Koldoba}  \&
  {Lovelace}}{{Romanova} et~al.}{2002}]{Romanova2002}
{Romanova} M.~M.,  {Ustyugova} G.~V.,  {Koldoba} A.~V.,   {Lovelace} R.~V.~E.,
  2002, \mn@doi [\apj] {10.1086/342464}, \href
  {https://ui.adsabs.harvard.edu/abs/2002ApJ...578..420R} {578, 420}

\bibitem[\protect\citeauthoryear{{Romanova}, {Ustyugova}, {Koldoba}  \&
  {Lovelace}}{{Romanova} et~al.}{2004}]{Romanova2004}
{Romanova} M.~M.,  {Ustyugova} G.~V.,  {Koldoba} A.~V.,   {Lovelace} R.~V.~E.,
  2004, \mn@doi [\apj] {10.1086/421867}, \href
  {https://ui.adsabs.harvard.edu/abs/2004ApJ...610..920R} {610, 920}

\bibitem[\protect\citeauthoryear{{Romanova}, {Kulkarni}  \&
  {Lovelace}}{{Romanova} et~al.}{2008}]{RomanovaKulkarni2008}
{Romanova} M.~M.,  {Kulkarni} A.~K.,   {Lovelace} R. V.~E.,  2008, \mn@doi
  [\apjl] {10.1086/527298}, \href
  {https://ui.adsabs.harvard.edu/abs/2008ApJ...673L.171R} {673, L171}

\bibitem[\protect\citeauthoryear{{Romanova}, {Ustyugova}, {Koldoba}  \&
  {Lovelace}}{{Romanova} et~al.}{2012}]{Romanova2012}
{Romanova} M.~M.,  {Ustyugova} G.~V.,  {Koldoba} A.~V.,   {Lovelace} R.~V.~E.,
  2012, \mn@doi [\mnras] {10.1111/j.1365-2966.2011.20055.x}, \href
  {https://ui.adsabs.harvard.edu/abs/2012MNRAS.421...63R} {421, 63}

\bibitem[\protect\citeauthoryear{{Romanova}, {Koldoba}, {Ustyugova}, {Blinova},
  {Lai}  \& {Lovelace}}{{Romanova} et~al.}{2021}]{RomanovaKoldobaEtAl2021}
{Romanova} M.~M.,  {Koldoba} A.~V.,  {Ustyugova} G.~V.,  {Blinova} A.~A.,
  {Lai} D.,   {Lovelace} R.~V.~E.,  2021, \mn@doi [\mnras]
  {10.1093/mnras/stab1724}, \href
  {https://ui.adsabs.harvard.edu/abs/2021MNRAS.506..372R} {506, 372}

\bibitem[\protect\citeauthoryear{{Salmi}, {N{\"a}ttil{\"a}}  \&
  {Poutanen}}{{Salmi} et~al.}{2018}]{Salmi18}
{Salmi} T.,  {N{\"a}ttil{\"a}} J.,   {Poutanen} J.,  2018, \mn@doi [\aap]
  {10.1051/0004-6361/201833348}, \href
  {https://ui.adsabs.harvard.edu/abs/2018A&A...618A.161S} {618, A161}

\bibitem[\protect\citeauthoryear{Shi, Stone  \& Huang}{Shi
  et~al.}{2016}]{ShiStoneEtAl2016}
Shi J.-M.,  Stone J.~M.,   Huang C.~X.,  2016, \mn@doi [Monthly Notices of the
  Royal Astronomical Society] {10.1093/mnras/stv2815}, 456, 2273

\bibitem[\protect\citeauthoryear{{Spitkovsky}}{{Spitkovsky}}{2006}]{Spitkovsky2006}
{Spitkovsky} A.,  2006, \mn@doi [\apjl] {10.1086/507518}, \href
  {https://ui.adsabs.harvard.edu/abs/2006ApJ...648L..51S} {648, L51}

\bibitem[\protect\citeauthoryear{Staubert et~al.,}{Staubert
  et~al.}{2019}]{StaubertTrumperEtAl2019}
Staubert R.,  et~al., 2019, \mn@doi [Astronomy and Astrophysics]
  {10.1051/0004-6361/201834479}, 622, A61

\bibitem[\protect\citeauthoryear{Stone \& Balbus}{Stone \&
  Balbus}{1996}]{StoneBalbus1996}
Stone J.~M.,  Balbus S.~A.,  1996, \mn@doi [The Astrophysical Journal]
  {10.1086/177328}, 464, 364

\bibitem[\protect\citeauthoryear{{Sur}, {Haskell}  \& {Kuhn}}{{Sur}
  et~al.}{2020}]{Sur2020}
{Sur} A.,  {Haskell} B.,   {Kuhn} E.,  2020, \mn@doi [\mnras]
  {10.1093/mnras/staa1212}, \href
  {https://ui.adsabs.harvard.edu/abs/2020MNRAS.495.1360S} {495, 1360}

\bibitem[\protect\citeauthoryear{{Suvorov} \& {Melatos}}{{Suvorov} \&
  {Melatos}}{2020}]{Suvorov2020}
{Suvorov} A.~G.,  {Melatos} A.,  2020, \mn@doi [\mnras]
  {10.1093/mnras/staa3132}, \href
  {https://ui.adsabs.harvard.edu/abs/2020MNRAS.499.3243S} {499, 3243}

\bibitem[\protect\citeauthoryear{Takasao, Tomida, Iwasaki  \& Suzuki}{Takasao
  et~al.}{2018}]{TakasaoTomidaEtAl2018}
Takasao S.,  Tomida K.,  Iwasaki K.,   Suzuki T.~K.,  2018, \mn@doi [The
  Astrophysical Journal] {10.3847/1538-4357/aab5b3}, 857, 4

\bibitem[\protect\citeauthoryear{Tchekhovskoy, McKinney  \&
  Narayan}{Tchekhovskoy et~al.}{2008}]{TchekhovskoyMcKinney2008}
Tchekhovskoy A.,  McKinney J.~C.,   Narayan R.,  2008, \mn@doi [Monthly Notices
  of the Royal Astronomical Society] {10.1111/j.1365-2966.2008.13425.x}, 388,
  551

\bibitem[\protect\citeauthoryear{{Tchekhovskoy}, {Spitkovsky}  \&
  {Li}}{{Tchekhovskoy} et~al.}{2013}]{Tchekhovskoy2013}
{Tchekhovskoy} A.,  {Spitkovsky} A.,   {Li} J.~G.,  2013, \mn@doi [\mnras]
  {10.1093/mnrasl/slt076}, \href
  {https://ui.adsabs.harvard.edu/abs/2013MNRAS.435L...1T} {435, L1}

\bibitem[\protect\citeauthoryear{Tchekhovskoy, Philippov  \&
  Spitkovsky}{Tchekhovskoy et~al.}{2015}]{TchekhovskoyPhilippov2015}
Tchekhovskoy A.,  Philippov A.,   Spitkovsky A.,  2015, preprint (\mn@eprint
  {arXiv} {1503.01467})

\bibitem[\protect\citeauthoryear{{Uzdensky}, {K{\"o}nigl}  \&
  {Litwin}}{{Uzdensky} et~al.}{2002a}]{Uzdensky2002a}
{Uzdensky} D.~A.,  {K{\"o}nigl} A.,   {Litwin} C.,  2002a, \mn@doi [\apj]
  {10.1086/324720}, \href
  {https://ui.adsabs.harvard.edu/abs/2002ApJ...565.1191U} {565, 1191}

\bibitem[\protect\citeauthoryear{{Uzdensky}, {K{\"o}nigl}  \&
  {Litwin}}{{Uzdensky} et~al.}{2002b}]{Uzdensky2002b}
{Uzdensky} D.~A.,  {K{\"o}nigl} A.,   {Litwin} C.,  2002b, \mn@doi [\apj]
  {10.1086/324724}, \href
  {https://ui.adsabs.harvard.edu/abs/2002ApJ...565.1205U} {565, 1205}

\bibitem[\protect\citeauthoryear{Vlahakis}{Vlahakis}{2004}]{Vlahakis2004}
Vlahakis N.,  2004, \mn@doi [The Astrophysical Journal] {10.1086/379701}, 600,
  324

\bibitem[\protect\citeauthoryear{{Wasserman} \& {Shapiro}}{{Wasserman} \&
  {Shapiro}}{1983}]{Wasserman1983}
{Wasserman} I.,  {Shapiro} S.~L.,  1983, \mn@doi [\apj] {10.1086/160745}, \href
  {https://ui.adsabs.harvard.edu/abs/1983ApJ...265.1036W} {265, 1036}

\bibitem[\protect\citeauthoryear{Watts et~al.,}{Watts et~al.}{2016}]{Watts2016}
Watts A.~L.,  et~al., 2016, \mn@doi [Rev. Mod. Phys.]
  {10.1103/RevModPhys.88.021001}, 88, 021001

\bibitem[\protect\citeauthoryear{{Zanni} \& {Ferreira}}{{Zanni} \&
  {Ferreira}}{2009}]{Zanni2009}
{Zanni} C.,  {Ferreira} J.,  2009, \mn@doi [\aap]
  {10.1051/0004-6361/200912879}, \href
  {https://ui.adsabs.harvard.edu/abs/2009A&A...508.1117Z} {508, 1117}

\bibitem[\protect\citeauthoryear{{Zanni} \& {Ferreira}}{{Zanni} \&
  {Ferreira}}{2013}]{Zanni2013}
{Zanni} C.,  {Ferreira} J.,  2013, \mn@doi [\aap]
  {10.1051/0004-6361/201220168}, \href
  {https://ui.adsabs.harvard.edu/abs/2013A&A...550A..99Z} {550, A99}

\bibitem[\protect\citeauthoryear{{van den Eijnden}, {Degenaar}, {Russell},
  {Wijnands}, {Miller-Jones}, {Sivakoff}  \& {Hern{\'a}ndez Santisteban}}{{van
  den Eijnden} et~al.}{2018}]{Eijnden2018}
{van den Eijnden} J.,  {Degenaar} N.,  {Russell} T.~D.,  {Wijnands} R.,
  {Miller-Jones} J.~C.~A.,  {Sivakoff} G.~R.,   {Hern{\'a}ndez Santisteban}
  J.~V.,  2018, \mn@doi [\nat] {10.1038/s41586-018-0524-1}, \href
  {https://ui.adsabs.harvard.edu/abs/2018Natur.562..233V} {562, 233}

\bibitem[\protect\citeauthoryear{{van den Eijnden} et~al.,}{{van den Eijnden}
  et~al.}{2021}]{Eijnden2021}
{van den Eijnden} J.,  et~al., 2021, \mn@doi [\mnras] {10.1093/mnras/stab1995},
  \href {https://ui.adsabs.harvard.edu/abs/2021MNRAS.507.3899V} {507, 3899}

\bibitem[\protect\citeauthoryear{{van der Klis}}{{van der
  Klis}}{2006}]{Klis2006}
{van der Klis} M.,  2006, \mn@doi [Advances in Space Research]
  {10.1016/j.asr.2005.11.026}, \href
  {https://ui.adsabs.harvard.edu/abs/2006AdSpR..38.2675V} {38, 2675}

\makeatother
\end{thebibliography}


\appendix

\section{Variability of the mass accretion rate}\label{sec:mdot}

\begin{figure}
\includegraphics[width=8.5cm]{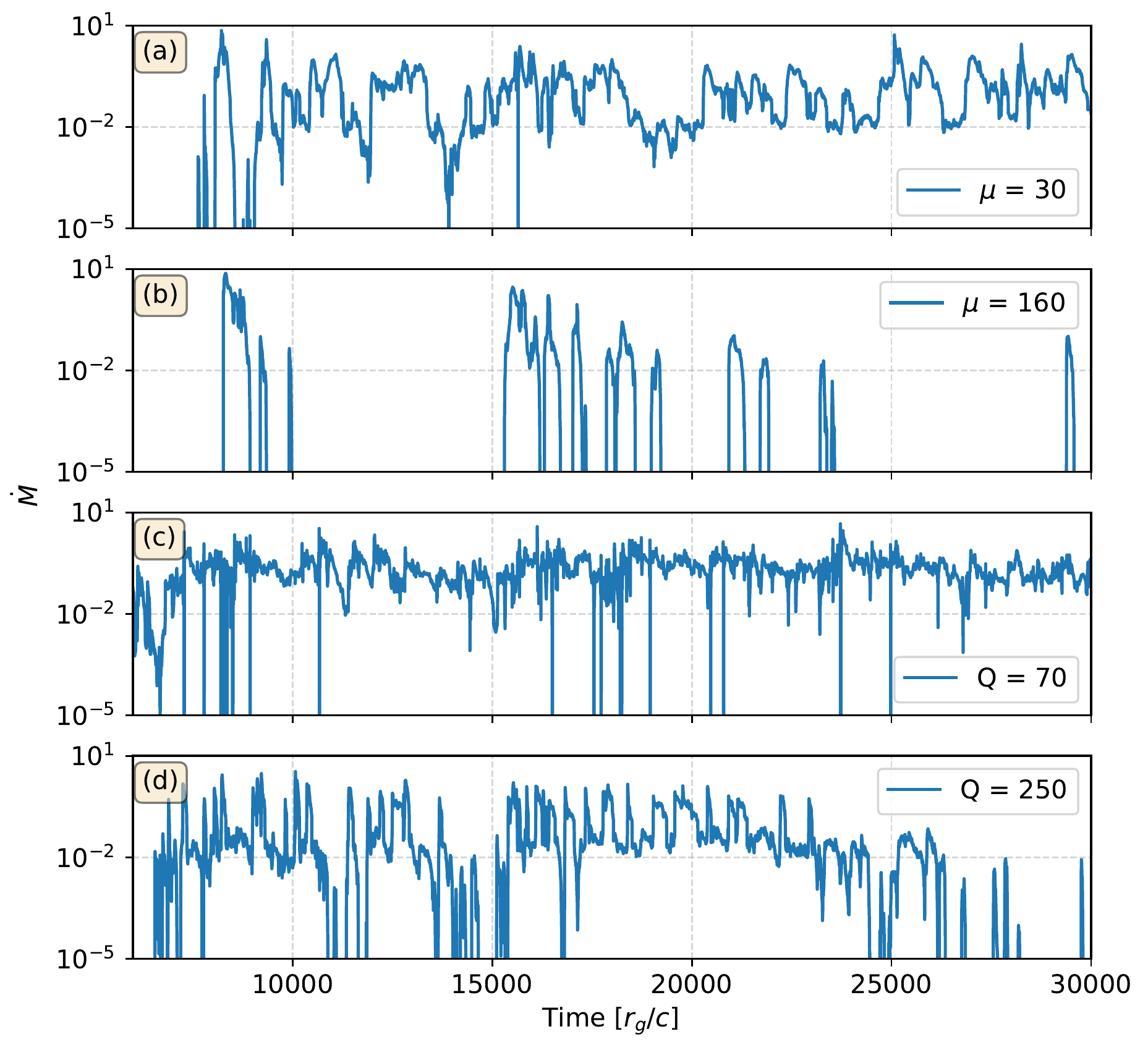}%
\caption{Mass accretion rate for stellar dipoles and quadrupoles rotating at $\Omega = 0.03$. (a) $\mu = 30$, channeled regime (Dipole)  (b) $\mu = 160$, Propeller regime (Dipole) (c) Q = 70 (Quadrupole) (d) Q = 250 (Quadrupole)}
\label{fig:mdot}
\end{figure}

Figure \ref{fig:mdot} shows the mass accretion rate in logarithmic scale for stellar dipoles and quadrupoles for a few selected cases with $\Omega = 0.03$ and different magnetic field strengths. The first panel (a) in Figure \ref{fig:mdot} is our fiducial channeled case with persistent columns. The second panel (b) shows the episodic behaviour in the propeller regime with some accretion streams occasionally falling into the star. 
The quadrupolar case (c) is also in the channeled regime and shows weaker variability than the $\mu=30$ dipole (also quantified by variability indices $c_{\rm v}=0.8605$ vs. $c_{\rm v}=1.4765$, c.f. Table \ref{tab:runparameters}).  Finally, the largest shown quadrupolar case (d) also occasionally enters into the propeller regime resulting in large drops in mass accretion ration.  
Common to all our models is that the variability increases with increasing magnetization (e.g. larger magnetospheric radii).  

\section{Asymmetry of the quadrupolar jets}\label{sec:asymm}
\begin{figure}
\includegraphics[width=8cm]{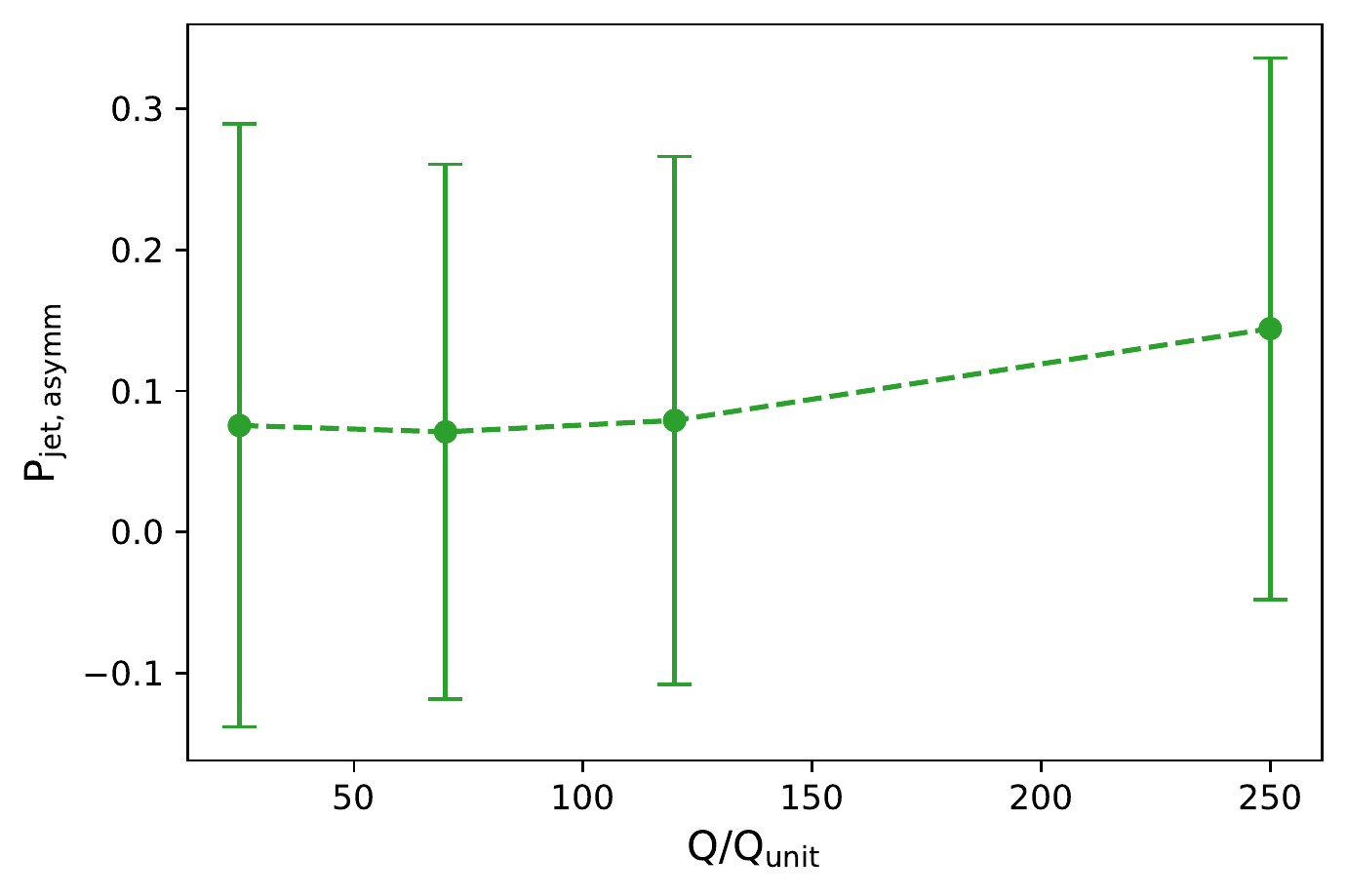}%
    \caption{Quadrupolar jet asymmetry as a function of surface polar magnetic field strength averaged over t $\in$ [10000,30000]$r_g/c$. The vertical bars show the standard deviation over the averaging interval.} %
    \label{fig:jet-assym-quad}
\end{figure}

The asymmetry parameter is defined as 
\begin{align}
P_{\rm jet, asymm} = \frac{P_{\rm jet,north} - P_{\rm jet,south}}{P_{\rm jet,north} + P_{\rm jet,south}} 
\end{align}                         
with $P_{\rm jet,north}$ and $P_{\rm jet,south}$ being the power in the upper and lower hemisphere respectively.  Hence $P_{\rm jet, asymm}=-1 , (+1)$ corresponds to a jet purely in the southern (northern) hemisphere and $P_{\rm jet, asymm}=0$ is the symmetric case.  
Figure \ref{fig:jet-assym-quad} shows the resulting asymmetry from our quadrupolar cases.  It is worth pointing out that jet asymmetry does not vanish even near the boundary layer regime (Q=25) and is only mildly dependent on the magnetization (i.e. we have $P_{\rm jet, asymm}=0.08 - 0.15$ essentially from the boundary-layer to the propeller regime).  

\section{Vector potentials and radial magnetic fields}

The dipolar, quadrupolar and quadrudipolar vector potentials are given as follows,
\begin{align}
A_{\phi}^{\rm dipole} = \frac{-3 \mu \sin^2\theta}{2M} \left[ x^2 {\rm log}\left(1-\frac{1}{x}\right) + x + \frac{1}{2}\right]
\end{align}
\begin{multline}
A_{\phi}^{\rm quadrupole} = \frac{5 Q \sin^2\theta \cos\theta}{M^2}\bigg[(3-4 x) x^2 \log \left(\frac{x-1}{x}\right)\\
    +(1-4 x) x+\frac{1}{6} \bigg]
\end{multline}
\begin{align}
A_{\phi}^{\rm quadrudipole} = A_{\phi}^{\rm dipole} + A_{\phi}^{\rm quadrupole}
\end{align}
where x = r/2M and $\mu$ and Q are the dipolar and quadrupolar moment respectively. The radial polar magnetic field strength at the stellar surface is given by,
\begin{align}
  B^r(\rstar,0)_{\rm dipole} &= 0.036\mu\\
  B^r(\rstar,0)_{\rm quadrupole} &= 0.013Q
\end{align}

\section{Temporal evolution of open and closed flux}\label{sec:fluxopening}
\begin{figure}
   \centering
    \includegraphics[width=8.5cm]{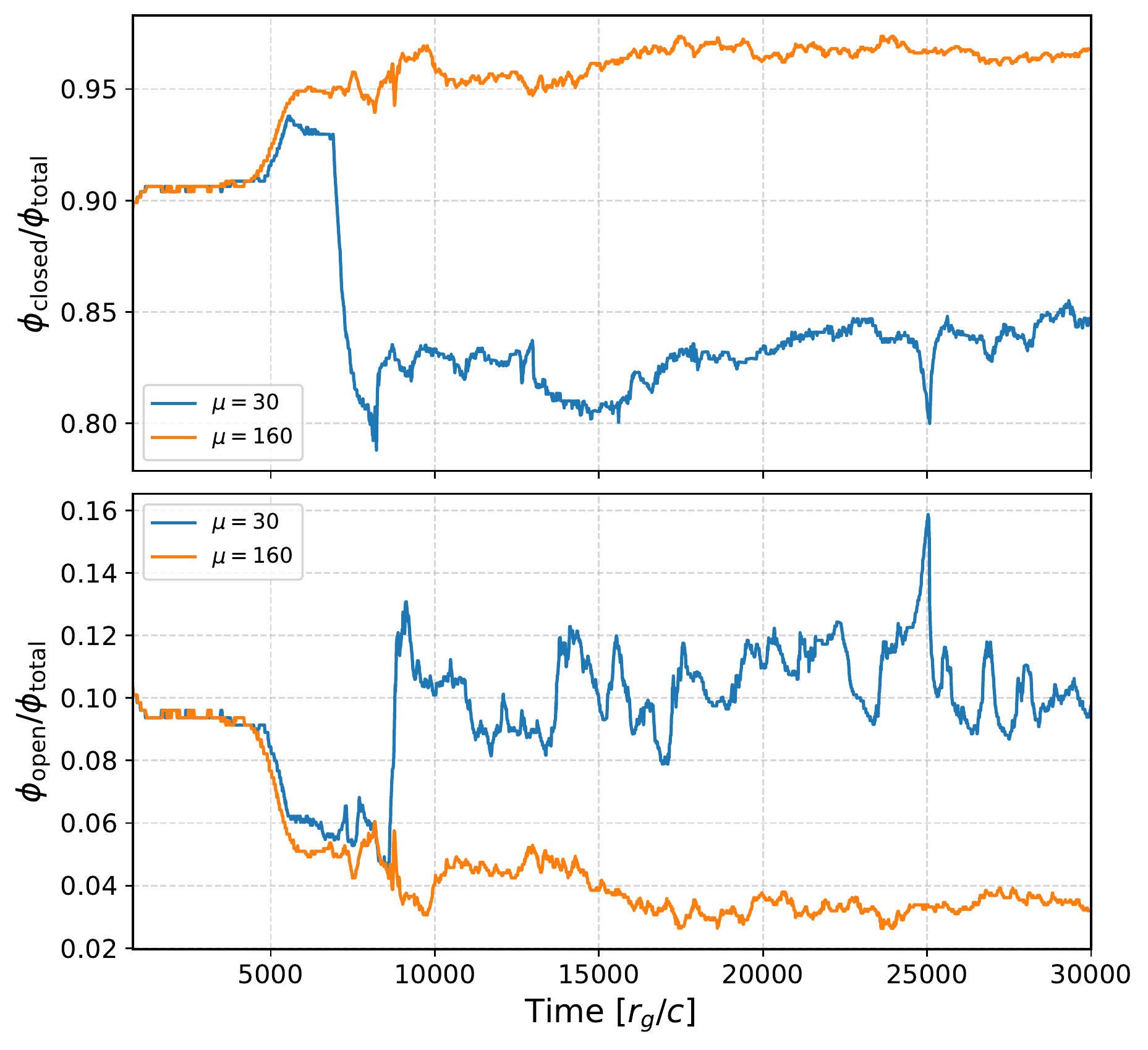}%
    \caption{Fraction of closed (upper panel) and open flux (lower panel) as a function of time for $\mu = 30$ and $\mu = 160$.}
    \label{fig:openfluxvstime}
\end{figure}
Figure \ref{fig:openfluxvstime} shows the evolution of open and closed flux for dipolar case with different surface polar field strength. For lower surface field strength, simillar to \citetalias{Parfrey2017}, the disk opens up initial closed stellar field lines leading to an increase in net open flux as shown in Figure \ref{fig:jet-power-di}. This was also observed in case of protostellar disk- magnetosphere interaction resulting in strong winds \citep{Ferreira2000}. For higher dipolar strength, the disk closes the initial open stellar flux and leads to an decrease and increase in open and closed flux respectively.


\bsp	
\label{lastpage}
\end{document}